\documentclass[preprint,12pt]{aastex}
\usepackage{times}
\usepackage{epsfig,graphicx}
\usepackage{txfonts}

\newcommand{\Msun}{$M_{\odot}$}
\newcommand{\Zsun}{$Z_{\odot}$}
\newcommand{\HI}{H{\small \sc I} }
\newcommand{\HIcom}{H{\small \sc I,} }

\newcommand{\kms}{km s$^{-1}$}
\newcommand{\h}{^h}
\newcommand{\m}{^m}
\newcommand{\s}{^s}
\newcommand{\dg}{^\circ}

\newcommand{\am}{'}

\newcommand{\as}{''}

\begin{document}

\title{ESO 381-47, an early-type galaxy with extended \HI and a star forming ring \\  \textnormal{\today}}
\author{Jennifer L. Donovan\altaffilmark{1}, Paolo Serra\altaffilmark{2}, J.~H. van Gorkom\altaffilmark{1}, S.~C. Trager\altaffilmark{3}, Tom Oosterloo\altaffilmark{2,3}, J.~E. Hibbard\altaffilmark{4}, Raffaella Morganti\altaffilmark{2,3}, David Schiminovich\altaffilmark{1}, J.~M. van der Hulst\altaffilmark{3}}
\altaffiltext{1}{Department of Astronomy, Columbia University, 550 West 120th St., Mail Code 5246, New York, NY 10027}
\altaffiltext{2}{ASTRON, Netherlands Institute for Radio Astronomy, Postbus 2, 7990~AA, Dwingeloo, The Netherlands}
\altaffiltext{3}{Kapteyn Astronomical Institute, University of Groningen, Postbus 800, NL-9700~AV Groningen, The Netherlands}
\altaffiltext{4}{NRAO, 520 Edgemont Rd., Charlottesville, VA 22903}




\begin{abstract}
ESO 381-47 is an early type galaxy with an extended \HI disk. GALEX and very deep optical images reveal a distinct stellar ring far outside the optical body with a diameter of $\sim$30 kpc, which has undergone recent star formation at 1.8 $\times$ 10$^{-4}$ \Msun~yr$^{-1}$~kpc$^{-2}$, consistent with other new results which detect low level star formation below the traditional Kennicutt relation in the outer parts of spiral galaxies. The morphology of this galaxy resembles the recently identified class of ultraviolet objects called extended ultraviolet disks, or XUV-disks. New \HI observations of this galaxy taken at the ATCA and in the CnB array at the VLA show that the cold gas lies in an extended (diameter $\sim$90 kpc) ring around the central S0 galaxy. The \HI data cube can be well modeled by a warped ring. The faint ionized gas in the inner parts of the galaxy is kinematically decoupled from the stars and instead appears to exhibit velocities consistent with the rotation of the \HI ring at larger radius. The peak of the stellar ring, as seen in the optical and UV, is slightly displaced to the inside relative to the peak of the \HI ring. We discuss the manner in which this offset could be caused by the propagation of a radial density wave through an existing stellar disk, perhaps triggered by a galaxy collision at the center of the disk, or possibly due to a spiral density wave set up at early times in a disk too hot to form a stellar bar. 
Gas accretion and resonance effects due to a bar which has since dissolved are also considered to explain the presence of the star forming ring seen in the GALEX and deep optical data.
\end{abstract}


\section{Introduction} 
\label{intro}

For a long time, the formation of stars far outside of the optical disks of galaxies went undetected. Ten years ago, pioneering work by \citet{Ferguson98} showed that stars do form in such extreme environments, and more recently GALEX images have shown that star formation outside of the optical bodies of galaxies is much more common than had been previously thought \citep{Thilker07}. Much effort has gone into correlating this phenomenon with gas surface density (e.g. \citealt{Lelievre00, Martin01, Thilker05, Oosterloo07_pro}). 
Studying the birthplaces of recently formed stars 
helps to elucidate the physics of star formation and whether these vary as a function of location in the galaxy. The analysis of the age and chemical composition of these stars can help to identify the triggers of star formation by, for example, establishing the importance of accretion from the intergalactic medium or interactions with other systems.

Using a local (d $<$ 40 Mpc) subset of the GALEX Atlas of Nearby Galaxies, \citet{Thilker07} established a system to classify a population of galaxies which exhibit such extra-disk ultraviolet emission -- the extended ultraviolet disk, or XUV-Disk population. Systems are classified as Type 1 XUV-disks if the outer disk exhibits structured ultraviolet emission beyond a traditional star formation threshold; they are called Type 2 if a large UV-bright region coincides with an optical low surface brightness (LSB) zone outside of the inner, older stellar populations. Type 1 XUV-disk features are found outside early- to late-type spirals while the Type 2 class primarily consists of late-types. 

In this paper, we present multiwavelength observations of the \HI-rich early-type galaxy ESO 381-47, which exhibits a ring in the ultraviolet. We discuss the galaxy in the framework of an early-type example of a Type 1 XUV-disk. We study the star forming ring and its origin by identifying potential triggers; also, as XUV-disks almost always exhibit extended \HI distributions, we study the specific nature of the cold gas and its relationship to the young stellar component. The morphology of the \HI in ESO 381-47 is that of an extended ring \citep{SerraThesis} with the UV-bright stellar ring near its inner edge. 


It is not uncommon for lenticular galaxies to have rings of \HI. In a study of 24 gas-rich S0, S0/a, and Sa systems, \citet{vanDriel91} find the presence of inner and outer rings of neutral gas to be more common than in their comparison sample of later type spirals. In the early type sample, roughly one-third of the galaxies are barred early type disks, and these systems always exhibit \HI rings or pseudorings (partial rings consisting of tightly wound spiral arms). To explain the origin and current state of the \HI, \citet{vanDriel91} suggest accretion from another galaxy or galaxies with sufficient subsequent time to allow the gas to settle into a regularly rotating structure. 
Alternatively, the gas may have been accreted from the intergalactic medium (IGM) \citep{Binney77, Keres05}). 

It is also not uncommon for early type galaxies to exhibit stellar rings; the most common cause for such phenomena is the presence of a bar. Complete rings or pseudorings can occur in resonances in the disk usually excited by a bar or the presence of some other nonaxisymmetric perturbation. Though they are much less common, there are plenty of examples of rings in nonbarred systems, created for instance by tidal forces from an interacting companion or by a bar which has since dissipated (for a comprehensive review, see \citealt{Buta96}) or even by a spiral density wave in the disk \citep{Rautiainen00}. 
Collisional ring systems, such as the Cartwheel, wherein an existing gas-rich disk is disrupted by a companion which falls through the center of the disk and creates a radially propagating density wave (e.g. \citealt{Appleton96}) are another possible explanation. 
Alternatively, \citet{Brook07} suggest that prograde mergers of gas-rich disk galaxies can also create ring galaxies. 

Stellar rings are often sites of ongoing star formation \citep{Buta96}, and indeed stars forming at large galactic radii are consistent with XUV-disk star formation as seen by \citet{Thilker07}. However, using \HI maps of five S0 galaxies from the thesis of \citet{vanDriel87}, \citet{Kennicutt89} finds that the surface density of cold gas is below the critical density for gravitational collapse, and therefore that star formation is not expected within the rings. Lenticular galaxies typically have high rotation velocities, corresponding to high critical densities for gravitational collapse, but measured gas surface densities far below that threshold, making star formation unlikely. 
Yet ESO 381-47 shows a UV ring. What triggered this star formation?

In this paper, we study the stellar component, gas-rich ring, and star formation outside of the optical disk associated with the early type galaxy ESO 381-47. In \S II, we provide a review of previous observations of this galaxy. In \S III, we describe our observations and the reduction of the data presented in this paper, the results of which are presented in \S IV. We construct a model to analyze the \HI structure of ESO 381-47, which is also described in this section. The characteristics of the star forming ring and possible formation scenarios for the \HI structure and UV ring are discussed in \S V. Our conclusions are summarized in \S VI. Throughout this paper, we use a distance of 61.2~Mpc to ESO 381-47, based on its recessional velocity of 4819 \kms, H$_{o}$=71 \kms~Mpc$^{-1}$, and correcting for Virgo, Great Attractor, and Shapley Supercluster infall (as listed in NED).

\section{Properties of ESO 381-47}
\label{properties}

Various studies in the literature have shown that the morphology of this galaxy is definitively early type, but there the agreement ends. 
ESO 381-47 was originally categorized as a lenticular, or S0, galaxy (RC2, \citealt{DeVaucouleurs76}; ESO/Uppsala, \citealt{Lauberts82}). A few years later, this classification was revised to E-S0 by \citet{Lauberts89} and a notation indicating that the galaxy is an ``S0 in a cluster" was added to the ESO/Uppsala Catalog. The updated RC3 catalog \citep{DeVaucouleurs91} revised the RC2 classification, listing ESO 381-47 as a doubtful S0, or S0?. The FLASH Redshift Survey \citep{Kaldare03}, drawn from the Hydra-Centaurus Catalogue \citep{Raychaudhury90} identifies it as E, as does the work by \citet{Dressler91} from visual inspection of the ESO (B) glass plates. Here, we present deep optical imaging which reveals the presence of a faint stellar disk, indicating that the system is indeed an S0. 
 
Originally, ESO 381-47 was detected in neutral hydrogen in the \HI Parkes All Sky Survey (HIPASS; \citealt{Barnes01}). The \HI was mapped with the Australia Telescope Compact Array (ATCA) by \citet{Oosterloo07}, showing the gas to be in a large, regularly rotating disk with an \HI mass of 6.3 x 10$^{9}$ \Msun~(adopting a distance of 61.2~Mpc, used throughout\footnote{From Table~\ref{table:generalproperties}, repeated here.}). In \citet{SerraThesis}, deeper ATCA observations showed the morphology of the disk to actually be in the shape of a ring; these observations are presented in this paper as well.

The optical properties for a sample of early-types with known \HI characteristics, including ESO 381-47, were published by \citet[hereafter S08]{Serra08}. They find an age for its central stellar population of 12.9 $^{+1.6}_{-1.2}$ Gyr. 
A more detailed study of its kinematics \citep{SerraThesis} shows the presence of a rotating stellar component on top of a non-rotating one, interpreted as evidence for a disk + bulge structure. We therefore consider ESO 381-47 to be an S0 galaxy.  

In this paper, a focused study of ESO 381-47 to follow up the studies by \citet{Oosterloo07} and S08, we discuss this system in the framework of an early type galaxy consisting of an old stellar population and surrounded by an extended ring of neutral gas. Here we present the much deeper ATCA observations of this galaxy also presented in \citet{SerraThesis} as well as deep and higher resolution Very Large Array (VLA) observations. The combined data show the large- and small-scale structure of the ring-shaped neutral hydrogen system of ESO 381-47.

ESO 381-47 is not an isolated system. The galaxy and its brightest neighbor, ESO 381-46, were already identified by \citet{Giuricin00} as a group, NOGG P2 695. In a study of galaxy systems within the Shapley Supercluster, \citet{Ragone06} identified a group, SSGC 22, in the field of ESO 381-47 with four members at a velocity similar to the optical velocity of our target, but the specific galaxy names are not listed. Our \HI observations (\S 3.3.1) reveal that ESO 381-47 is a member of a gas-rich group with at least four other galaxies, including ESO 381-46. All five galaxies are detected in our GALEX images. We also detect a sixth small companion galaxy, which we call ESO 381-47A, within the optical ring in our $V$, $R$, and UV imaging. 

ESO 381-47 has published magnitudes in the 2MASS All-Sky Extended Source Catalog (XSC) in the J, H, and K bands, but the ring is not detected in the NIR by 2MASS. 
No data is available in the FIR, as this galaxy does not appear in the IRAS Point Source Catalog, nor has it been observed by Spitzer. No X-ray counterpart is detected in the ROSAT Broad Band All-Sky Survey, down to a brightness limit of 0.1 cts s$^{-1}$. A summary of the galaxy's basic properties is given in Table~\ref{table:generalproperties}.

\begin{table}[t]
\small
\begin{center}
\centerline{Table~1: Optical and \HI properties of ESO 381-47}
\medskip
\begin{tabular}{lll}
\hline
\hline
RA (J2000) & 13$\h$01$\m$05.4$\s$ & a \\
Dec (J2000) & $-35\dg$36$\am$59.9$\as$ & a \\
Type & S0 & b, d \\
Distance & 61.2 Mpc & c, f \\
M$_{B}$ & $-20.15$ & b \\
$\sigma$ & 188 \kms & d \\
v$_{opt}$ & 4771 $\pm$ 51 \kms & e \\
\HI mass & 6.7 $\times$ 10$^{9}$ \Msun & f \\
v$_{\HI}$ & 4819 $\pm$ 7 \kms & f \\
\hline
\end{tabular}
\end{center}
\caption{$\it{References: a,}$ 2MASS XSC; $\it{b,}$ \citet{DeVaucouleurs91}; $\it{c,}$ distance derived from \HI measurement of 4819 \kms (this paper), corrected for Virgo, Great Attractor, and Shapley Supercluster infall (taken from NED) and assuming H$_{o}$=71 \kms~Mpc$^{-1}$; $\it{d,}$ S08; $\it{e,}$ \citet{Kaldare03}; $\it{f,}$ this paper. The quoted velocity dispersion is the central stellar velocity dispersion. The heliocentric \HI velocity is defined according to the optical definition, and we quote an uncertainty of the width of one channel, or 7 \kms, in v$_{\HI}$. 
\label{table:generalproperties}}
\end{table}

\section{Observations and data reduction}
\label{models}

The observations of ESO 381-47 presented in this paper include deep optical imaging performed at NTT/EMMI at La Silla, near-UV and far-UV imaging taken by GALEX, 1.4 GHz continuum imaging from the VLA, and combined \HI data from the ATCA and the VLA. \\

\subsection{Optical imaging}

Broadband images of ESO 381-47 in the $V$ and $R_C$ (hereafter simply $R$) bands were taken at the La Silla Observatory on 6 February 2005, using the red arm of EMMI (3800 -- 7020 $\AA$) on the NTT, operated by ESO. This instrument has a field of view of 22.7$\times$22.7 arcmin$^{2}$, and by binning 2$\times$2 pixels, a scale of 0.332 arcsec/pixel is achieved. The time on source for ESO 381-47 was 600 s in the $V$ band and 300 s in the $R$ band, during which the average seeing was 0.8$\as$. The data reduction was achieved using standard Python scripts and IRAF. Flat fielding was done using twilight sky-flats. Zeropoints were determined to magnitude-calibrate the data using multiple observations of three standard fields, each containing 4-7 bright Landolt standard stars; the magnitudes of these stars are given on the Johnson-Kron-Cousins scale in \citet{Landolt92}. A stellar mask was created for each band in order to keep stars in the field from contaminating measurements of the galaxy flux. 

The total $V$ magnitude and $R$ magnitude are measured inside a D$_{25}$ = 65.8$\as$ aperture \citep{DeVaucouleurs91} to be $V$=13.57 $\pm$ 0.03 and $R$=12.96 $\pm$ 0.04; the $V-R$ color is 0.61. Previously, the $R$ magnitude of ESO 381-47 was measured to be 13.01 by \citet{Lauberts89} inside a similar aperture, which is in good agreement with the measurement presented here. Correcting for Galactic extinction, we adjust our measurements by 0.18 mag in the $V$ band and 0.15 mag in the $R$ band, using the values listed in NED. Extinction-corrected values are $V$=13.39 $\pm$ 0.03, $R$=12.81 $\pm$ 0.04, and $V-R$=0.58.




\subsection{Ultraviolet observations}

We obtained GALEX data \citep{Martin05} on ESO 381-47 in the near-UV (NUV; 1750 -- 2800 $\AA$) and far-UV (FUV; 1350 -- 1750 $\AA$) bands over four orbits (total exposure time: 5980 seconds) and one orbit (total exposure time: 1630 seconds), respectively. 
The pixel size of the GALEX detectors is 1.5$\as$, and the FWHM of the NUV (FUV) image is 5.3$\as$ (4.2$\as$). The data reduction and calibration processes were performed by the GALEX pipeline, described in \citet{Morrissey07}, and we adopt NUV=20.08 and FUV=18.82 for the photometric zero points. 

\begin{figure}
\includegraphics[width=5in]{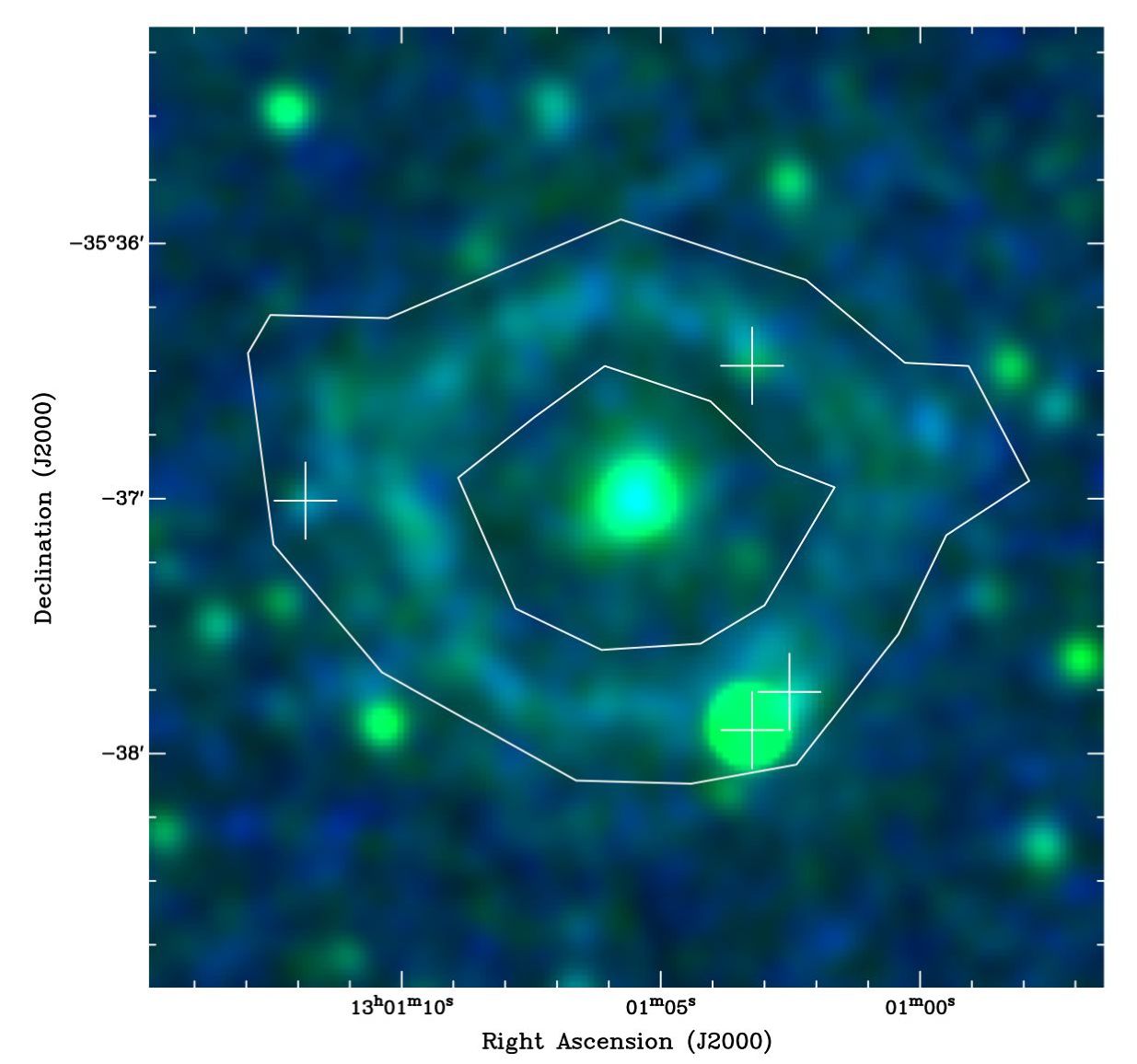}
\caption{GALEX image of NUV (green) and FUV (blue) data, shown smoothed to a resolution of 8.0$\as$. The integrated measurement of the ultraviolet emission in the ring was made from unsmoothed data between the two polygons shown. Crosses indicate the positions of stars and ESO 381-47A whose contaminating emission was subtracted from this measurement. \label{ring}}
\end{figure}

The data clearly show a ring in the ultraviolet. Both the NUV and FUV images were convolved with a Gaussian to a resolution of 8.0$\as$ 
in order to visually enhance the faint extended emission and facilitate our comparison to the \HI data. In Figure~\ref{ring}, we show the smoothed NUV (green) and FUV (blue) data, clearly showing the presence of the ultraviolet ring. For the purposes of magnitude measurements, we use unsmoothed, background-subtracted data. 
The ultraviolet properties of ESO 381-47 and the other four galaxies in the field are summarized in Table~\ref{sfrs}. For each galaxy, we measure flux in the unsmoothed images using ellipses defined at $m(AB)\sim27.4$ in the smoothed FUV images. The values quoted for the ring around ESO 381-47 are calculated by integrating all ultraviolet emission between two polygons, as shown in Figure~\ref{ring}. The polygons were chosen based on the smoothed data, which is also shown. Crosses in this image indicate the positions of stars (and ESO 381-47A, immediately northwest of the brightest star) between the polygons whose contaminating emission was not included in the ultraviolet measurements of the ring.

\begin{table}[t]
\small
\begin{center}
\centerline{Table~2: Derived Properties of Galaxies in the ESO 381-47 Field}
\medskip
\begin{tabular}{lccccc}
\hline
\hline
 & NUV (AB) & FUV (AB) & SFR$_{NUV}$ (10$^{-2}$ \Msun~yr$^{-1}$) & S$_{1.4GHz}$ (mJy) & SFR$_{1.4GHz}$ (10$^{-2}$ \Msun~yr$^{-1}$) \\
\hline
ESO 381-47 & 19.3 $\pm$ 0.2 & 21.4 $\pm$ 0.4 & (7.0) & $<$0.33 & $<$8.9 \\
--, ring & 18.2 $\pm$ 0.5 & 19.3 $\pm$ 1.0 & 19 & $<$2.3 & $<$61 \\
ESO 381-47A & 21.6 $\pm$ 0.06 & 22.8 $\pm$ 0.1 & 0.82 & $<$0.33 & $<$8.9 \\
2MFGC 10338 & 18.3 $\pm$ 0.2 & 18.9 $\pm$ 0.4 & 17 & 2.1 $\pm$ 0.33 & 56 \\
ESO 381-43N & 19.1 $\pm$ 0.1 & 19.8 $\pm$ 0.3 & 8.0 & $<$0.39 & $<$10 \\
ESO 381-43 & 18.7 $\pm$ 0.2 & 19.3 $\pm$ 0.4 & 12 & $<$0.39 & $<$10 \\
ESO 381-46 & 16.8 $\pm$ 0.2 & 17.4 $\pm$ 0.5 & 65 & 2.7 $\pm$ 0.44 & 71 \\
\hline
\end{tabular}
\end{center}
\caption{Observed and derived properties of the gas-rich galaxies in the field of ESO 381-47. NUV and FUV magnitudes are listed in AB units, and star formation rates are derived from the NUV counts according to the relation in \citet{KennicuttReview}. The published photometric precision of GALEX is 0.05 mag in the FUV band and 0.03 mag in the NUV band for point sources \citep{Morrissey07}; quoted uncertainties scale this precision by the aperture used to make each measurement. The NUV star formation rate for ESO 381-47 is listed in parentheses since the UV emission is expected to be from old, hot stars, not a young stellar population. Radio continuum flux densities (or upper limits) at 1.4~GHz are also listed (see \S 3.3.2) as well as the star formation rates corresponding to these values \citep{Yun01}. \label{sfrs}}
\end{table}

\begin{table}[t]
\small
\begin{center}
\centerline{Table~3: Summary of Radio Observations}
\medskip
\begin{tabular}{cccc}
\hline
\hline
& ATCA & VLA & Combined \\ 

\hline

Pointing center, RA & 13$\h$01$\m$05.399$\s$ & 13$\h$01$\m$05.400$\s$ & 13$\h$01$\m$05.400$\s$ \\
Pointing center, Dec & $-35\dg36\am59.996\as$ & $-35\dg36\am60.000\as$ & $-35\dg36\am60.000\as$ \\
Configuration & 750A, EW352 & CnB & ... \\
Longest baseline & 750~m, 350~m & 3~km & ... \\
Shortest baseline & 75~m & 35~m & ... \\
Time on source & 2x12h, 12h & 18h & ... \\
Bandwidth & 9.375 MHz & 3.125 MHz & 3.125 MHz \\
Velocity resolution & 7 \kms & 5 \kms & 7 \kms \\
Central velocity$^{a}$ & 4771 \kms & 4771 \kms & 4771 \kms \\
Beam size & 117.43 x 62.68 & 18.15 x 13.19 & 57.71 x 34.10 \\
Robust parameter$^{b}$ & 5 & 1 & 5 \\
RMS in one channel (mJy beam$^{-1}$) & 1.6 & 0.8 & 0.85 \\
1$\sigma$ N$_{\HI}$ (cm$^{-2}$) & 1.7 $\times$ 10$^{18}$ & 1.8 $\times$ 10$^{19}$ & 3.3 $\times$ 10$^{18}$ \\
\HI mass of ESO 381-47 ($\times$ 10$^{9}$ \Msun) & 6.3 & 1.9 & 6.7 \\

\hline

\end{tabular}
\end{center}
\caption{Parameters of the radio observations and data reduction for the ATCA and VLA data as well as for the combined (VLA+ATCA) cube. The ATCA bandwidth listed is that used in the analysis; the total bandwidth of the observations is 16~MHz. \HI mass is calculated assuming $\it{d}$ = 61.2~Mpc. $\it{a:}$ Heliocentric radial velocity for each set of observations using the optical definition. $\it{b:}$ A robust parameter of 5 corresponds to purely natural weighting. \label{radio}}
\end{table}

\begin{table}[t]
\small
\begin{center}
\centerline{Table~4: \HI Properties of Galaxies in the ESO 381-47 Field}
\medskip
\begin{tabular}{lcccccc}
\hline
\hline
 & RA$_{J2000}$ & Dec$_{J2000}$ & M$_{B}$ & v$_{\HI}$ (\kms) & \HI (10$^{8}$ \Msun) & $\Delta$v (\kms) \\
\hline
ESO 381-47 & 13$\h$01$\m$05.4$\s$ & $-35\dg$37$\am$00$\as$ & $-20.15$ & ... & ... & ... \\
--, ring & ... & ... & ... & 4819 & 67 $\pm$ 8 & 105 \\
ESO 381-47A & 13$\h$01$\m$2.5$\s$ & $-35\dg$37$\am$46$\as$ & ... & ... & ... & ... \\
2MFGC 10338 & 13$\h$01$\m$45.9$\s$ & $-35\dg$35$\am$30$\as$ & ... & 4990 & 2.1 $\pm$ 0.7 & 169 \\
ESO 381-43N & 13$\h$00$\m$45$\s$ & $-35\dg$44$\am$42$\as$ & ... & 4686 & 0.93 $\pm$ 0.3 & 74 \\
ESO 381-43 & 13$\h$00$\m$39.4$\s$ & $-35\dg$48$\am$43$\as$ & $-17.16$ & 4720 & 1.7 $\pm$ 0.5 & 147 \\
ESO 381-46 & 13$\h$01$\m$01.9$\s$ & $-35\dg$55$\am$54$\as$ & $-19.20$ & 4705 & 18 $\pm$ 3 & 266 \\
\hline
\end{tabular}
\end{center}
\caption{Observational properties of the five \HI-detected galaxies in the field of ESO 381-47. M$_{B}$ is calculated from the value listed in the RC3 catalog \citep{DeVaucouleurs91} for ESO 381-47 and the ESO-LV Catalog \citep{Lauberts89} for ESO 381-43 and ESO 381-46; B magnitudes are unpublished for the other systems. \HI heliocentric velocities are determined using the optical definition of velocity. The \HI mass and $\Delta$v are calculated from the neutral hydrogen observations presented in this paper. Quoted values for $\Delta$v indicate the velocity range over which \HI is detected at the position of each galaxy. The \HI emission associated with ESO 381-47 is listed under the ``ring" label, but it is associated with both the galaxy itself and ESO 381-47A as well. \label{opthiprops}}
\end{table}

\subsection{Radio observations}

\subsubsection{HI}

Observations of the \HI transition at the optical position of ESO 381-47 were made at the ATCA\footnote{The Australia Telescope Compact Array is part of the Australia Telescope which is funded by the Commonwealth of Australia for operation as a National Facility managed by CSIRO.} in two different configurations on 7 March and 12 April, 2004 to study the large-scale cold-gas properties of this galaxy. 
To further explore the nature of the star forming ring found by GALEX, we obtained higher resolution \HI imaging in the C north-B hybrid configuration at the VLA\footnote{The Very Large Array (VLA) is operated by the National Radio Astronomy Observatory, which is a facility of the National Science Foundation (NSF), operated under cooperative agreement by Associated Universities, Inc.} on 29 September, 6 October, and 8 October 2006. The hybrid arrays at the VLA utilize a longer northern arm to produce a rounder beam for low-declination sources. The observational parameters for each set of observations are listed in Table~\ref{radio}. We also list the \HI masses derived from each set of observations. Comparing these masses, it is obvious that the ATCA recovers more flux due to its better surface brightness sensitivity, though applying a taper in the $\it{uv}$ plane to the VLA data and imaging with natural weighting recovers more than half of the missing flux.

\begin{figure}
\includegraphics[width=5in]{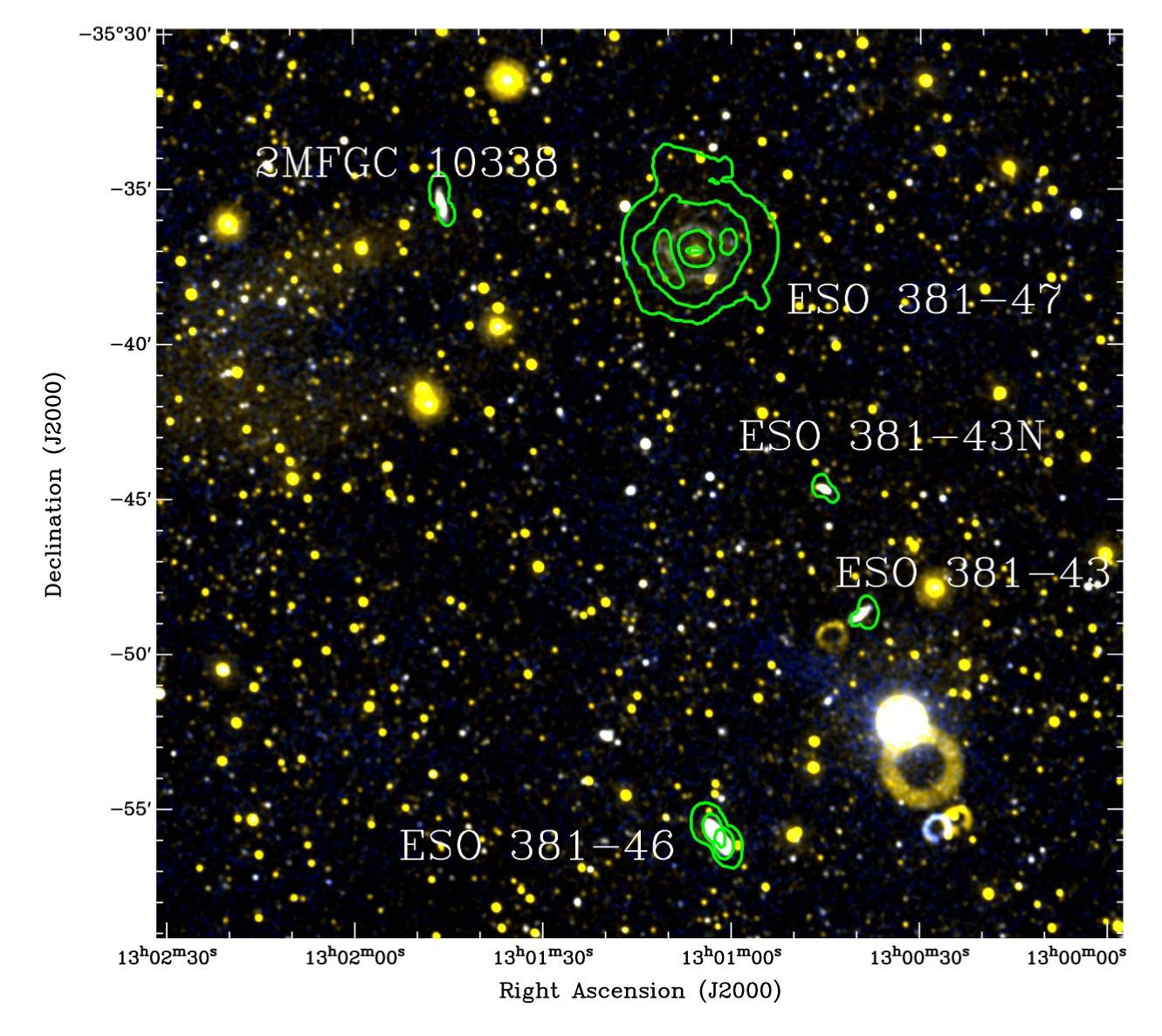}
\caption{\HI contours overlaid on smoothed GALEX data, where NUV is yellow and FUV is blue. Green \HI contours of the combined VLA+ATCA data are shown at levels=(2.8, 14, 25)$\times$10$^{19}$ cm$^{-2}$.  \label{uvHI}}
\end{figure}

We regridded the VLA data in frequency space to match the frequency resolution of the ATCA data and combined the ATCA and VLA cubes channel-by-channel in the $\it{uv}$-plane using AIPS. The line emission is derived by subtracting the continuum in the $\it{uv}$-plane using a linear fit to line-free channels. 
We image the \HI by utilizing a Gaussian taper of size 13.75 k$\lambda$ and a robust parameter of value 5 (natural weighting) to optimize sensitivity. Each channel of the combined image was then run through CLEAN for typically $\sim$100-400 iterations in boxes drawn around the emission regions. 
The rms of the final combined cube in each channel is 0.85 mJy beam$^{-1}$. 
An image of the total \HI emission of ESO 381-47 is made by summing over the full resolution cube after blanking pixels that are below a 2$\sigma$ threshold in a smoothed cube. To create the mask, the cube is convolved with a smoothing kernel with dimensions of 8 spatial pixels and 6 velocity channels. This yields a flux of 7.6 Jy \kms and an \HI mass of 6.7$\times$10$^{9}$ \Msun~for ESO 381-47. This map is shown in Figure~\ref{uvHI} overlaid on our ultraviolet image of the field. \HI masses were calculated according to the relation

\begin{displaymath}
$$ M$_{HI}$ (M$_{\odot}$) = 2.36 $\times$ 10$^{5}$ d$_{Mpc}^{2}$ S$_{HI}$, $$
\end{displaymath}

\noindent where $\it{d}$ is the distance measured in Mpc and S$_{HI}$ is the integrated \HI flux in Jy \kms. Primary beam corrections were approximated by taking an average of the VLA and ATCA primary beam corrections at the angular distance of each galaxy from the center of the pointing; no correction is necessary for ESO 381-47.

Typical uncertainties in measurements of the systemic velocity and velocity widths are less than the width of a channel, 7 \kms. The quoted uncertainties in the \HI masses are based on the rms noise in the channel maps 
and taking into account the number of velocity channels and number of independent synthesized beams that contribute to the summed signal. The rms noise goes up at large distances from the field center due to the primary beam correction, and this is taken into account in the quoted uncertainties as well. \HI masses, velocities, and velocity widths are listed in Table~\ref{opthiprops}. 

\subsubsection{1.4 GHz continuum}

\begin{figure}
\includegraphics[width=6in]{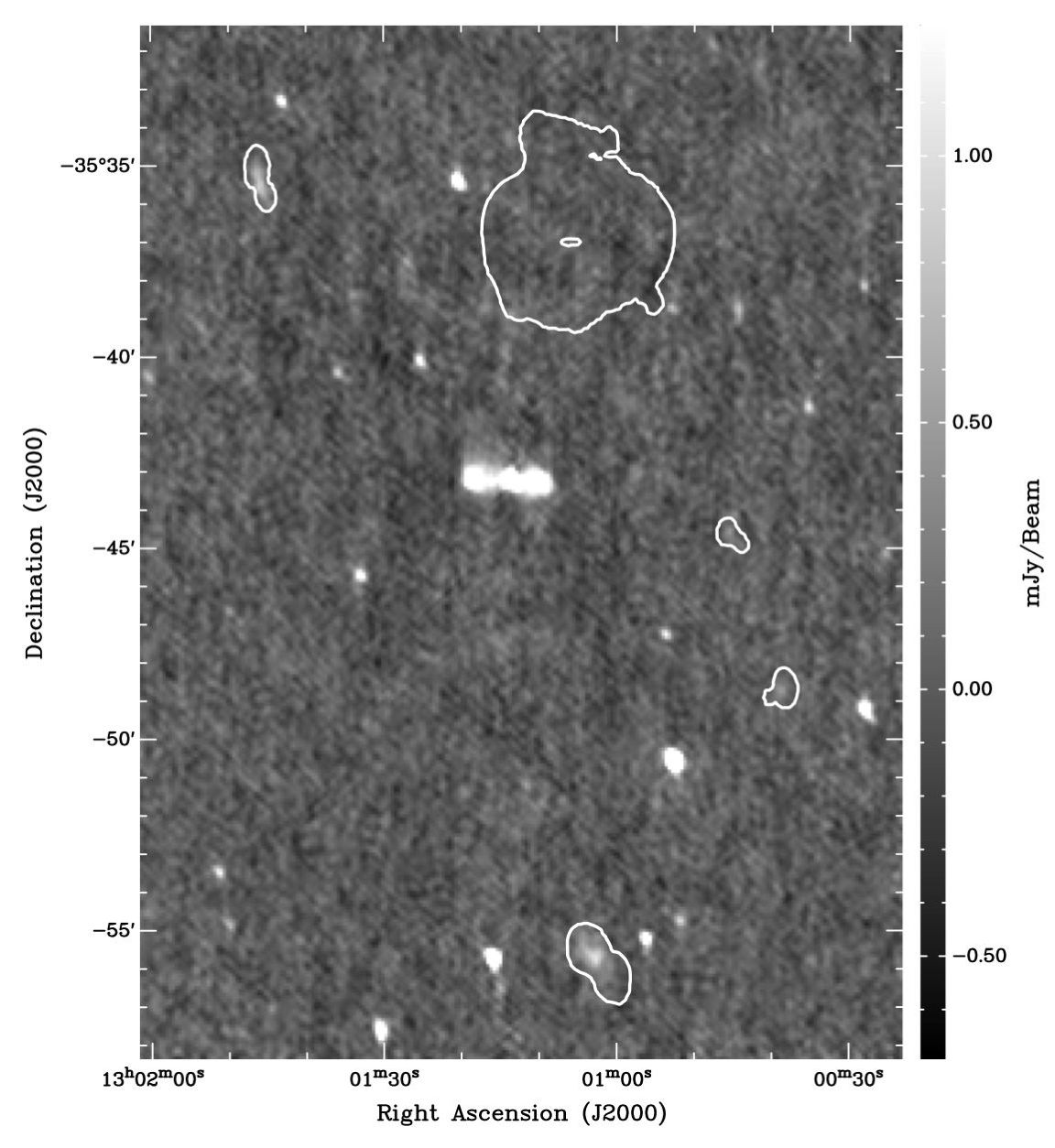}
\caption{1.4 GHz continuum image of the field made from our high resolution VLA data from all channels that are line-free for the \HI ring. The lowest \HI contour of the combined VLA+ATCA data, 2.8$\times$10$^{19}$ cm$^{-2}$, is overlaid to guide the eye. We detect 2MFGC 10338 and ESO 381-46 at 1.4 GHz; the continuum level within the other three galaxies is consistent with noise. Continuum measurements for those two galaxies are made from separate, appropriately line-free cubes. There appear to be faint detections of ESO 381-43N and ESO 381-43, but these are due to unsubtracted line emission in the displayed data.
\label{continuum}}
\end{figure}

The VLA-only data were used to make an image of the continuum emission in the ESO 381-47 field, shown in Figure~\ref{continuum}, using all channels which were line-free for the ring to search for continuum emission within the \HI structure. Detections and upper limits for the entire field are listed in Table~\ref{sfrs}. 

No emission is detected at the location of ESO 381-47 down to a 3$\sigma$ upper limit of 0.33 mJy beam$^{-1}$, significantly lower than the previous upper limit of 1.2 mJy quoted by \citet{Oosterloo07} for the central galaxy. Neither do we detect emission associated with the ultraviolet ring, and we quote a 3$\sigma$ upper limit integrated over its area. For each of the other upper limits in Table~\ref{sfrs}, we quote the limit on the emission contained in one beam. 

Two of the galaxies in the group, ESO 381-46 and 2MFGC 10338, exhibit extended continuum emission; see Table~\ref{sfrs} for these values. These measurements were made from slightly different data cubes, as the specific line channels corresponding to each galaxy needed to be removed. The rms of the resulting images is 0.13 mJy beam$^{-1}$ (slightly higher since most channels had to be removed due to the the large \HI velocity extent of the system) and 0.11 mJy beam$^{-1}$, respectively. Quoted uncertainties are the appropriate 1$\sigma$ values, scaled by the number of (VLA) beams subtended by ESO 381-46 and 2MFGC 10338. We also use the ESO 381-46 cube to search for continuum in ESO 381-43N and ESO 381-43 since they exhibit \HI at similar velocities; no continuum emission is detected at these positions, and so we quote the 3$\sigma$ upper limit of 0.39 mJy for these galaxies (again, assuming that the emission is contained in one beam).

The star formation rates corresponding to the detections and upper limits are listed in Table~\ref{sfrs}, calculated from the relation in \citet{Yun01} assuming that the emission is entirely due to young stars. This relation relies on the FIR star formation density from \citet{Buat96}, which is based on an initial mass function (IMF) with a high mass power law index of $-2.5$. 

We also detect a double lobed source at $\alpha$=13$\h$01$\m$13.4$\s$, $\delta$=-35$\dg$43$\am$12$\as$, with an integrated flux density of 53.8 mJy. The source is coincident with a ROSAT X-ray detection of source J130113.3-354315 from the All-Sky Survey Bright Source Catalog. 

\section{Results}
\label{results}


\subsection{Stellar morphology and age}

\begin{figure}
\begin{center}
\includegraphics[width=4in]{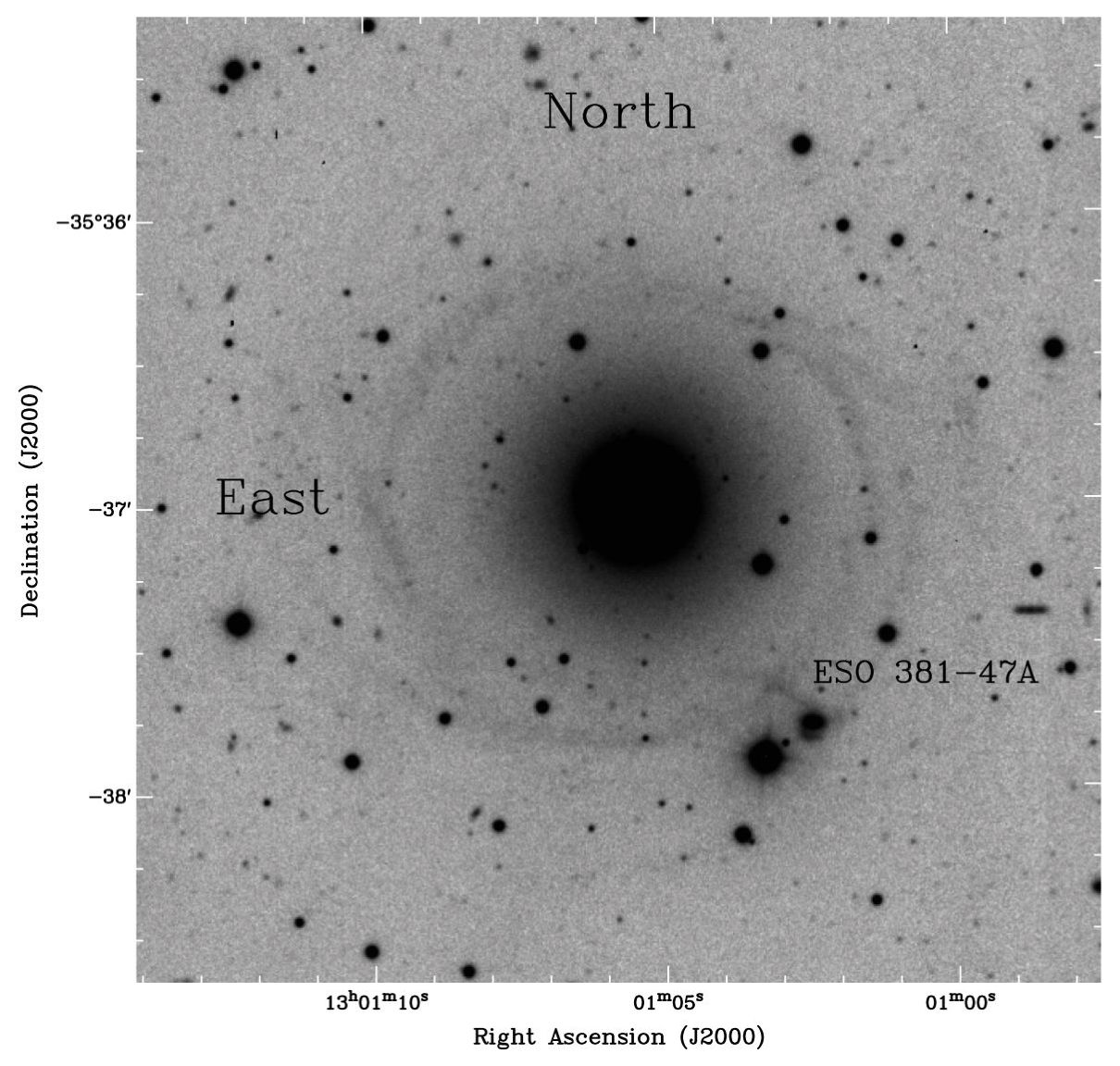} \\
\includegraphics[width=4in]{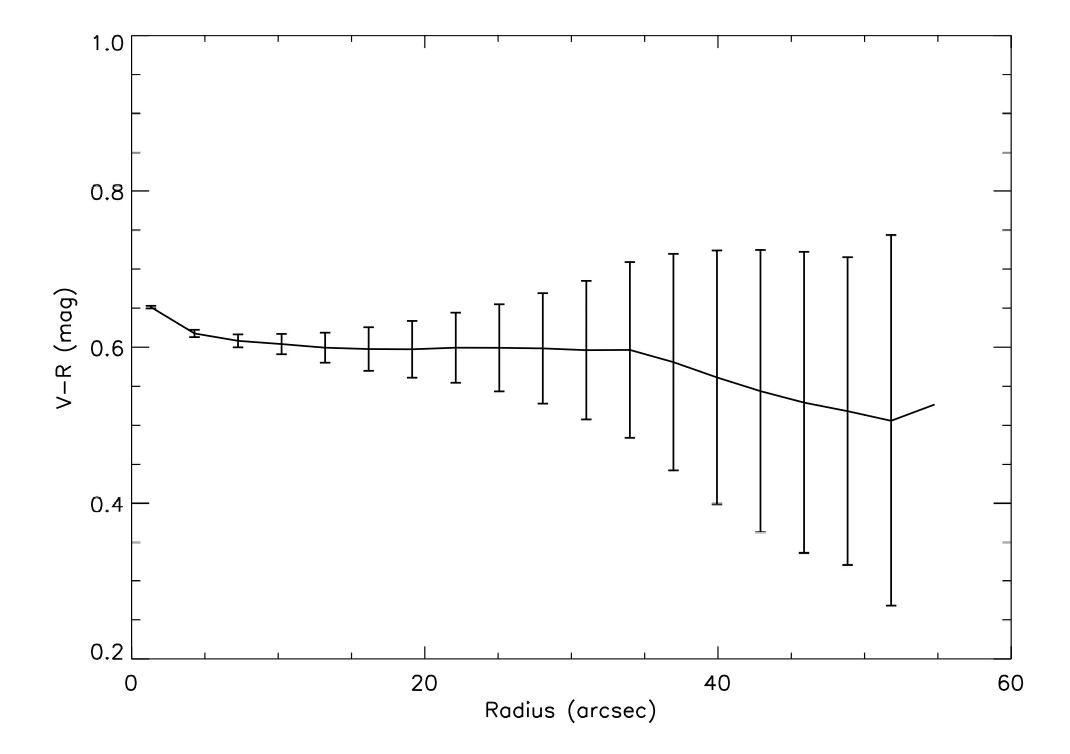}
\caption{$V$-band image and $V-R$ azimuthally averaged color profile of ESO 381-47. North and east are indicated. Note the small, disturbed companion galaxy ESO 381-47A, located southwest of ESO 381-47. The color profile is shown in 3$\as$ bins. The radius of the ring is $\sim$45~$\as$. \label{vband}}
\end{center}
\end{figure}

The $V$-band image of ESO 381-47 is shown in Figure~\ref{vband}. Low level, tightly wound spiral arms form an obvious ring structure, as they appear almost symmetrically around the galaxy at a projected radius of $\sim$45$\as$ ($\sim$14 kpc at the adopted distance of ESO 381-47). ESO 381-47A, a small companion galaxy, appears on the southwestern edge of the ring (projected radius $\sim$17 kpc), immediately northwest of a bright foreground star. It appears to be quite disturbed and in the process of being stripped of its stellar material, as a tail extends from its northeastern side, pointed directly toward the center of ESO 381-47, which suggests that it is undergoing a strong interaction with the central galaxy. (This interaction will be discussed further in \S 4.3.) Also in Figure~\ref{vband}, we show the azimuthally averaged $V-R$ color profile as a function of radius; in this profile, the light from the stars and companion galaxy have been removed. The suggestion of a flattening profile with subsequent steepening beyond a radius of 30$\as$ is interesting, but clearly the photometric errors are too large to conclude such evolution in the radial profile. 

\begin{figure}
\includegraphics[width=5in]{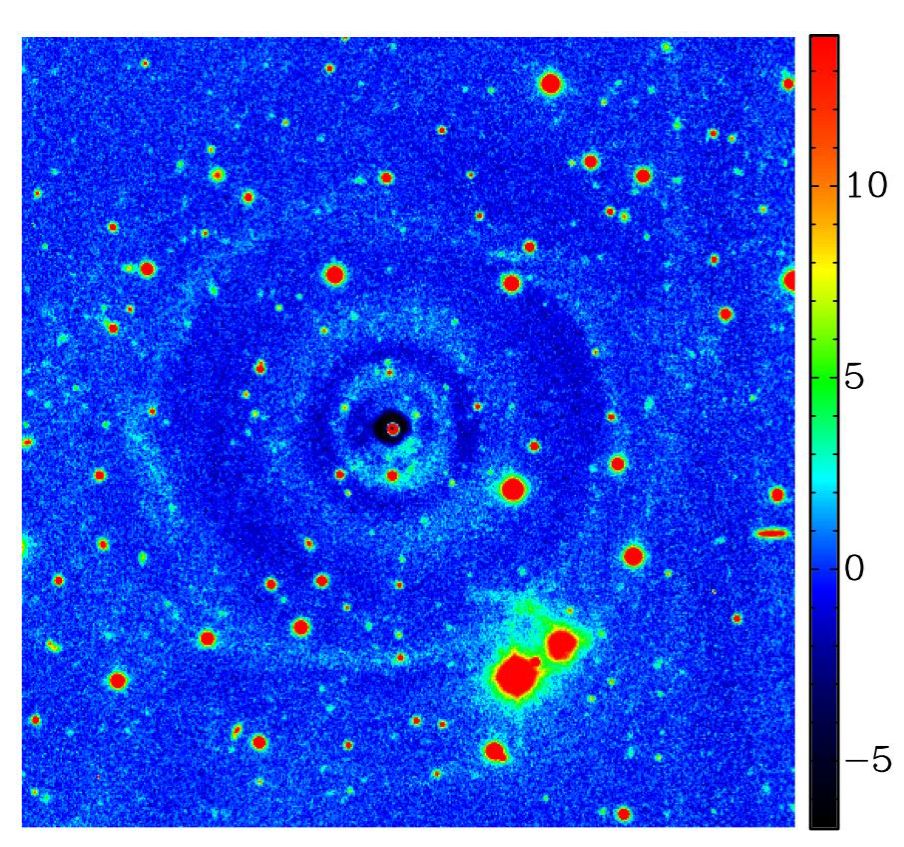}
\caption{Residual image of ESO 381-47 after subtraction of the bulge+disk model from our $R$-band image. The image is shown in units of the noise image. \label{rband} }
\end{figure}

We fit a bulge+disk morphology to the $R$-band image of ESO 381-47 using GALFIT \citep{Peng02}. The best fitting model consists of a Sersic bulge 
and an exponential disk with zero inclination angle. We model both the bulge and disk to be perfectly circular, since relaxing this constraint does not appreciably change the results. In addition, the centers of the bulge and disk are fixed to be at the same location. The residual of the $R$-band image minus the model is shown in Figure~\ref{rband}. There is no sign of the presence of an unsubtracted bar in the residuals. 

The presence of a stellar disk within the central 8-10\as was already known from \citet{SerraThesis}. The residuals at this radius may be caused by a non-exponential component of the disk which is not fit by our model. However, the inner ``rings" of residual emission are non-uniform, suggesting the presence of stellar features whose roles are not governed by the larger bulge or disk. The dips between the ``rings" in the residual image, where the residual counts drop below zero, may be significant as well and will be further discussed in \S 5.3.4. The outermost ring seen in the subtracted image coincides with the ring seen in the ultraviolet. 

\begin{figure}
\includegraphics[height=3in]{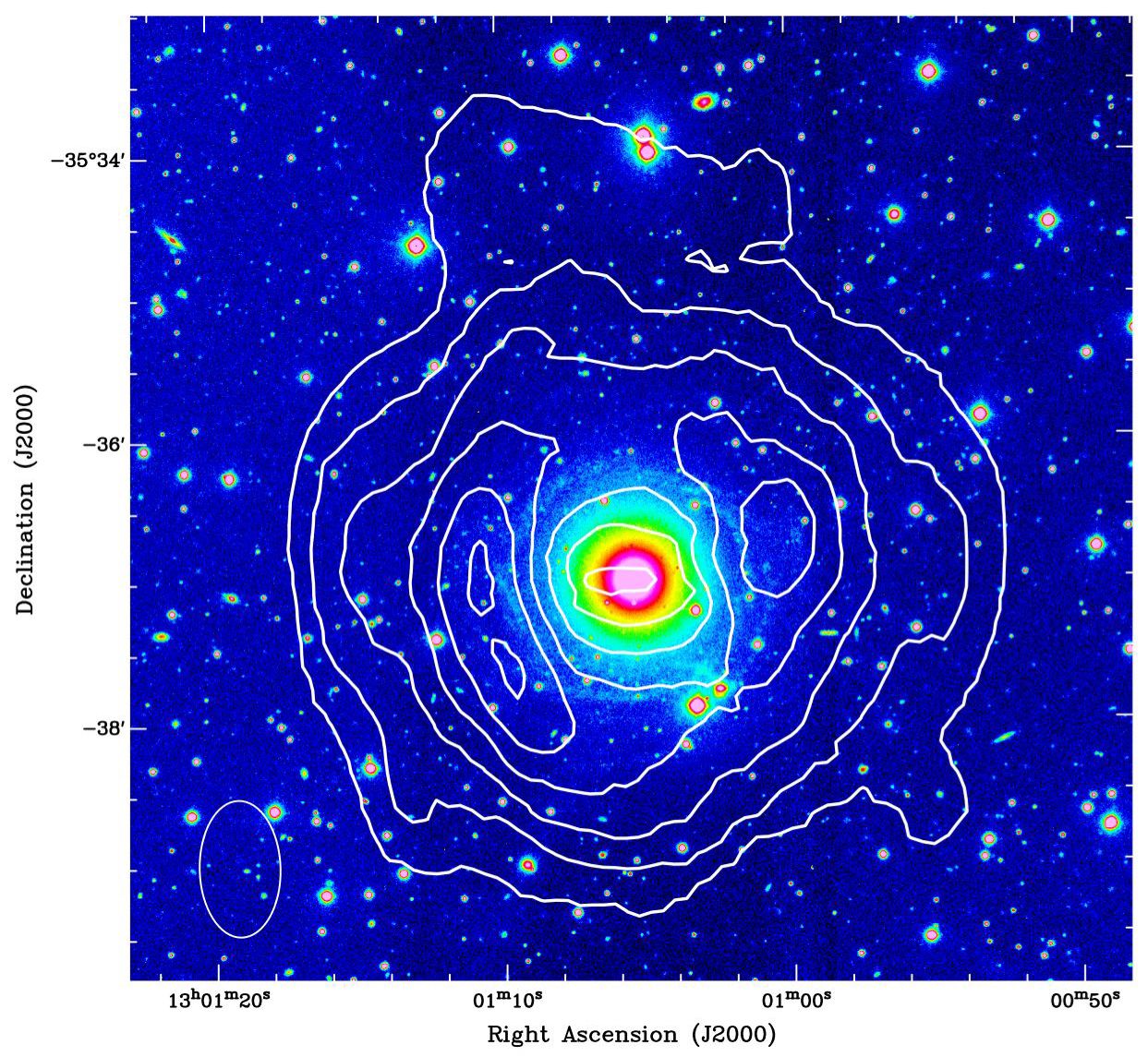}
\includegraphics[height=3in]{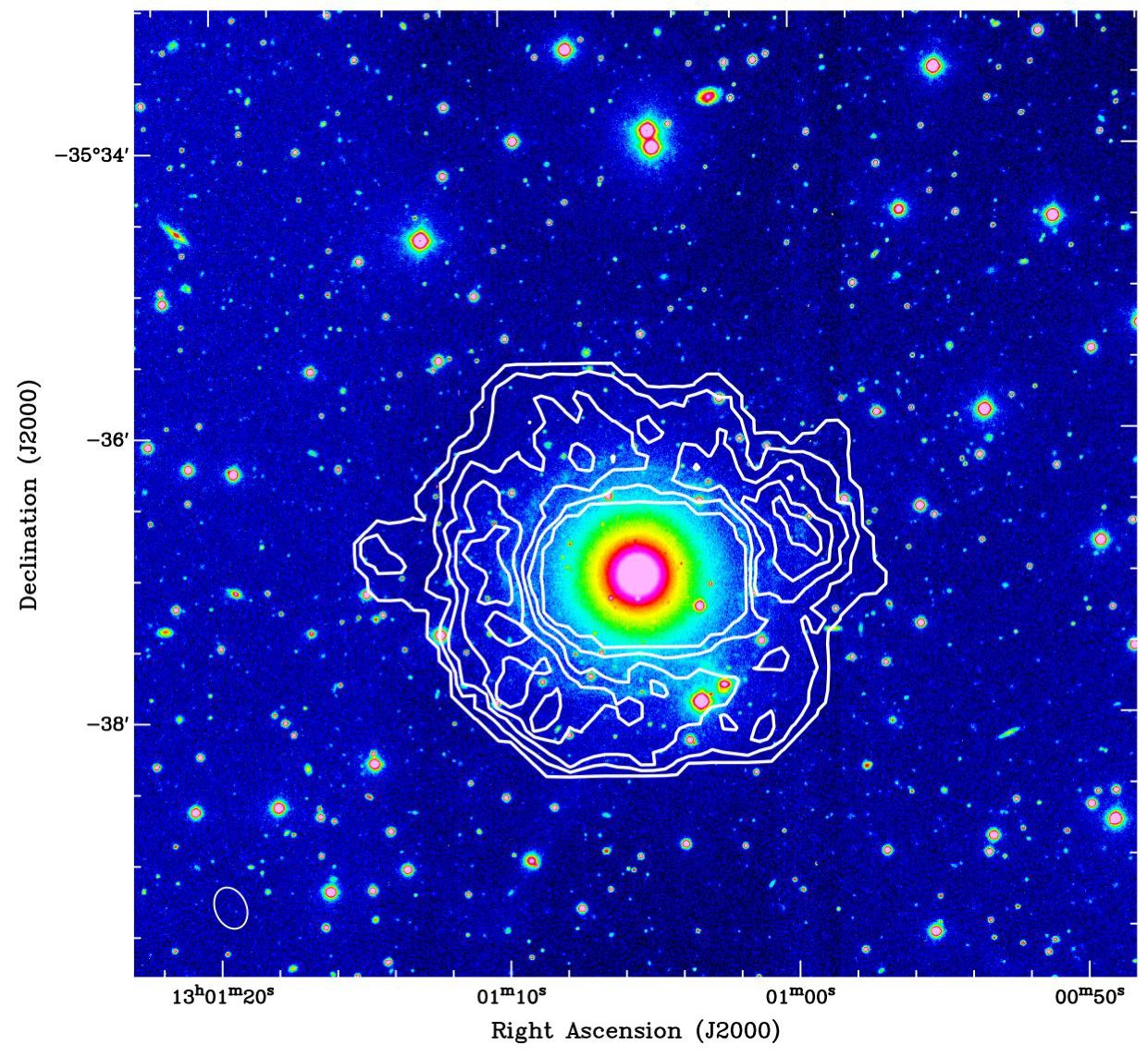} \\
\includegraphics[height=3in]{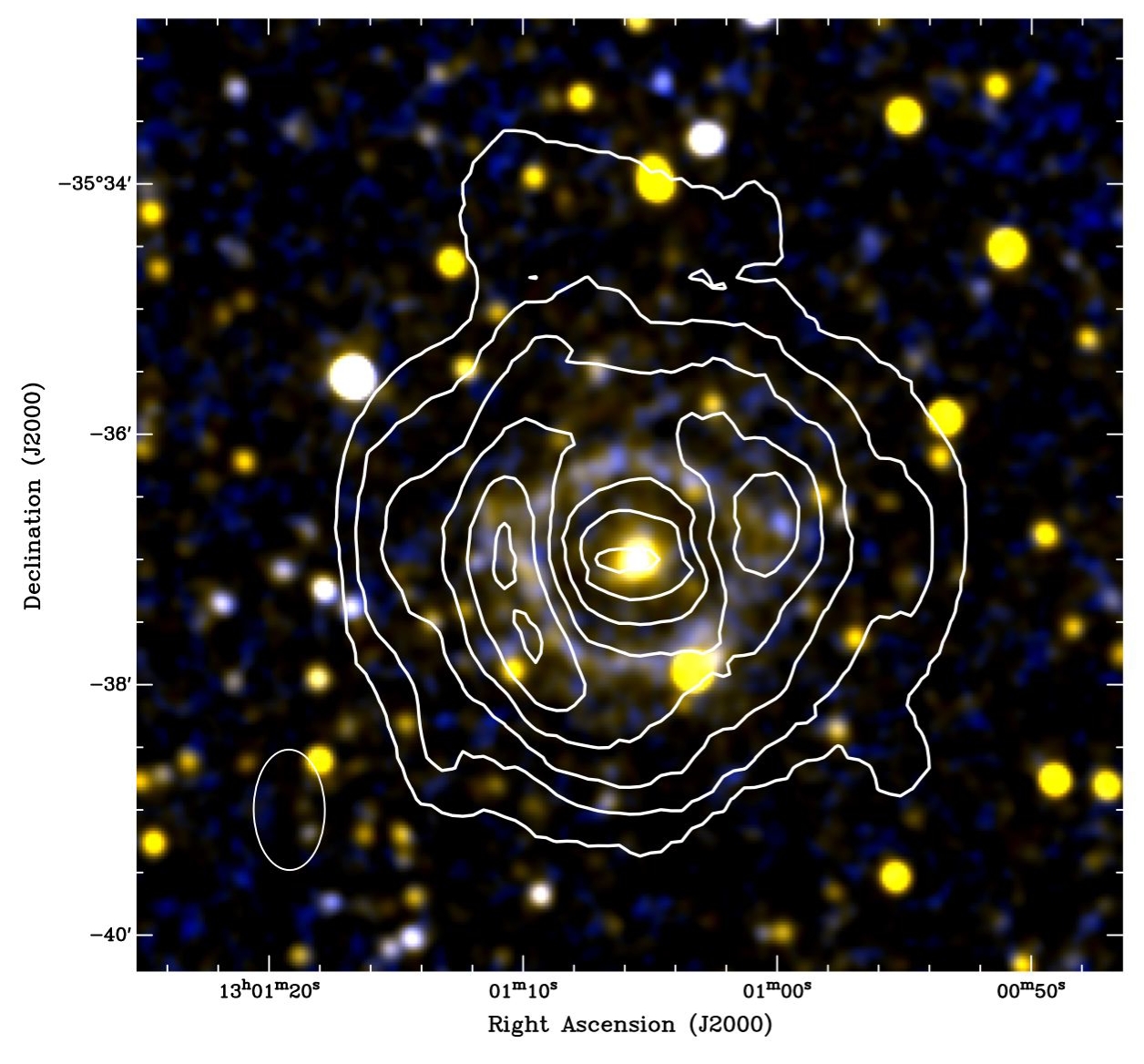}
\includegraphics[height=3in]{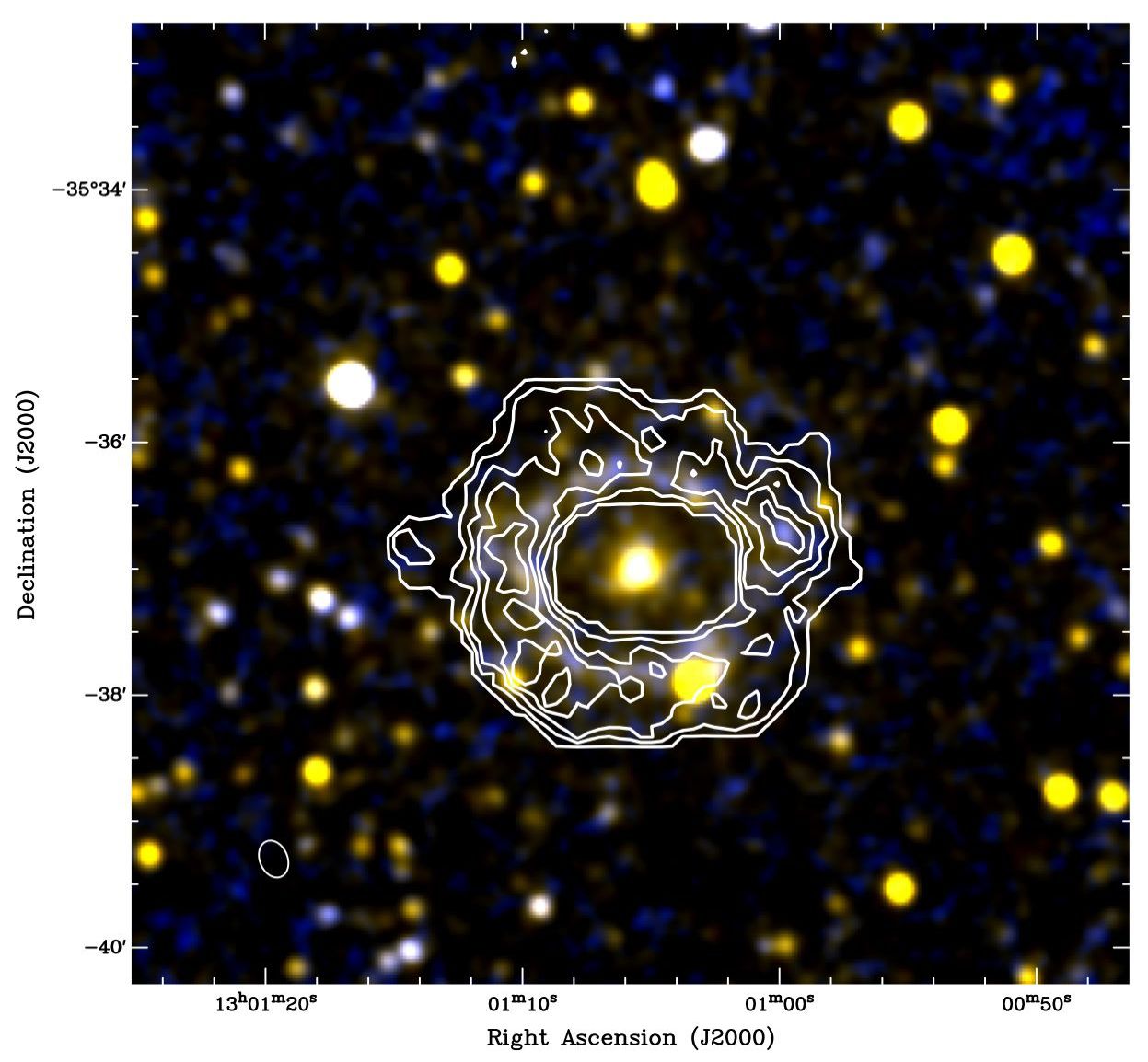} \\
\caption{\HI data ($\it{left}$, VLA+ATCA; $\it{right}$, VLA-only) overlaid on our $V$-band (top) and NUV+FUV (bottom) images of ESO 381-47. \HI contours are (2.8, 8.4, 14, 20, 25, 31)$\times$10$^{19}$ cm$^{-2}$ for the combined data and (2.3, 12, 23, 35, 46)$\times$ 10$^{19}$ cm$^{-2}$ for the VLA-only data. The \HI beam for each dataset is shown in the lower left of each panel for reference. \label{dss}}
\end{figure}

The integrated \HI emission from the combined VLA+ATCA data and VLA-only data is shown overlaid in contours on the smoothed NUV+FUV and V-band images in Figure~\ref{dss}. 
ESO 381-47 is very bright in the NUV and is surrounded by a comparatively faint but complete star forming ring. Both the central galaxy and ring are present but fainter in the FUV. There is a relative peak in FUV emission which corresponds to the position of ESO 381-47A (as seen in the optical data), while the NUV peak in that area is dominated by the foreground star. The other four \HI-detected galaxies are all detected in the FUV, as seen in Figure~\ref{uvHI}. The magnitudes derived for each system are listed in Table~\ref{sfrs}; these galaxies in the environment of ESO 381-47 will be discussed in \S 4.5.

For the main body of ESO 381-47, the UV emission is most likely due to old, very hot stars since comparing our NUV, FUV, and $R$ band measurements (converted to SDSS $r$ via \citealt{Jester05}) to Figure~3 in \citet{Yi05} shows that ESO 381-47 is consistent with their ``UV-weak" population, which is considered to be composed mainly of old stars. This is consistent with the age analysis by S08, which finds no young stellar component at the center of the galaxy. 

\begin{figure}
\includegraphics[height=2.5in]{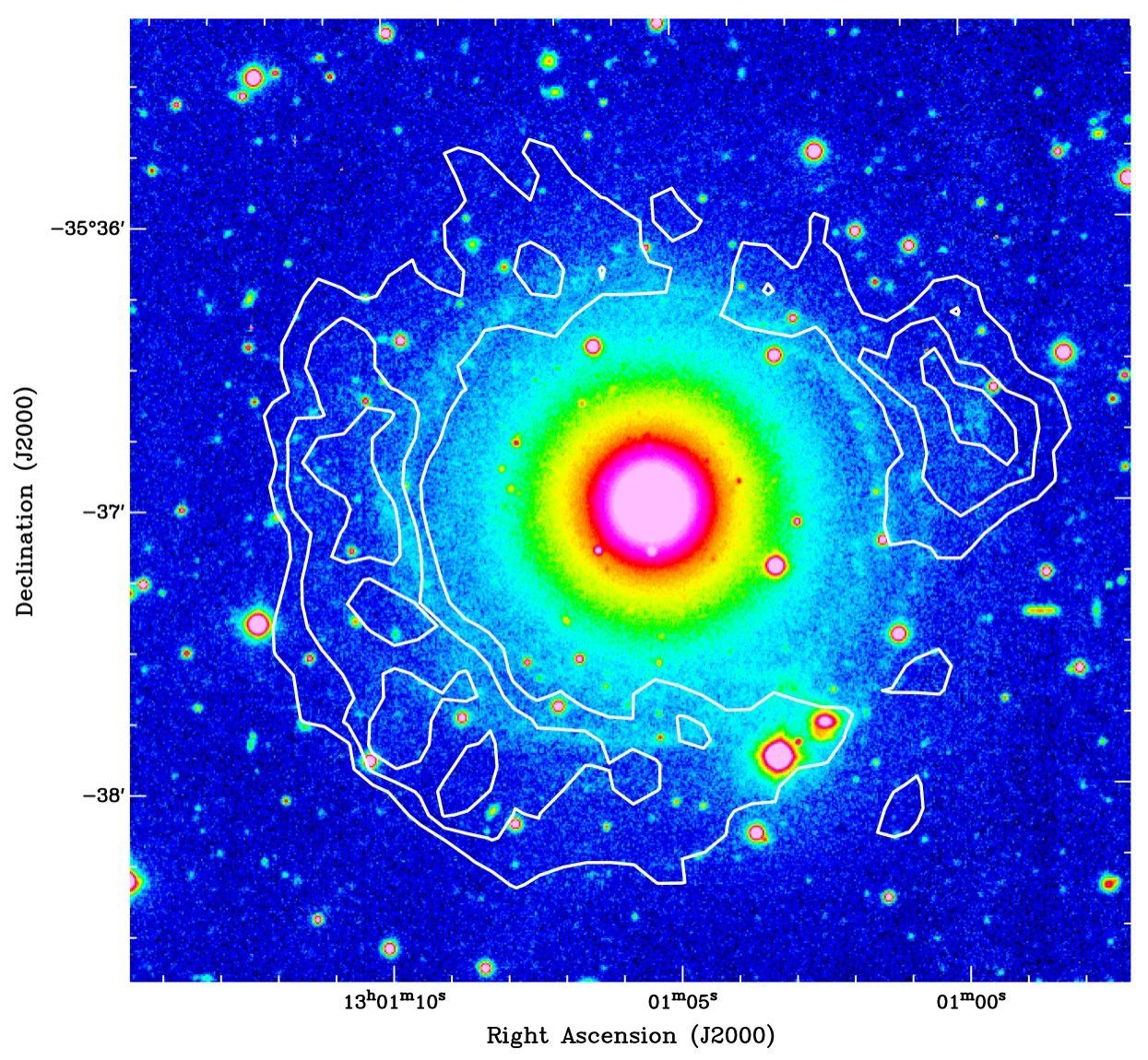} \\
\includegraphics[height=2.5in]{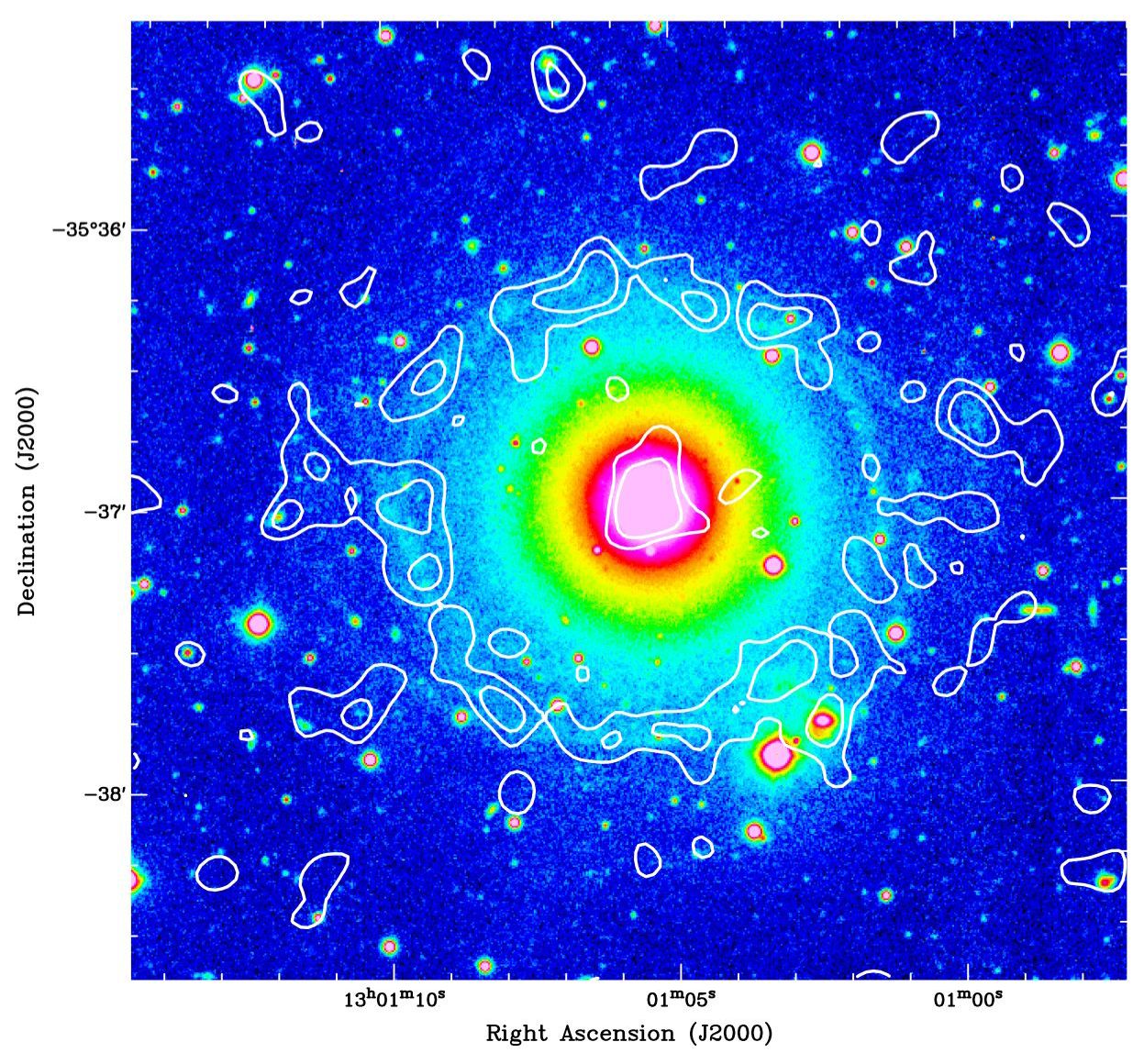} \\
\includegraphics[height=2.5in]{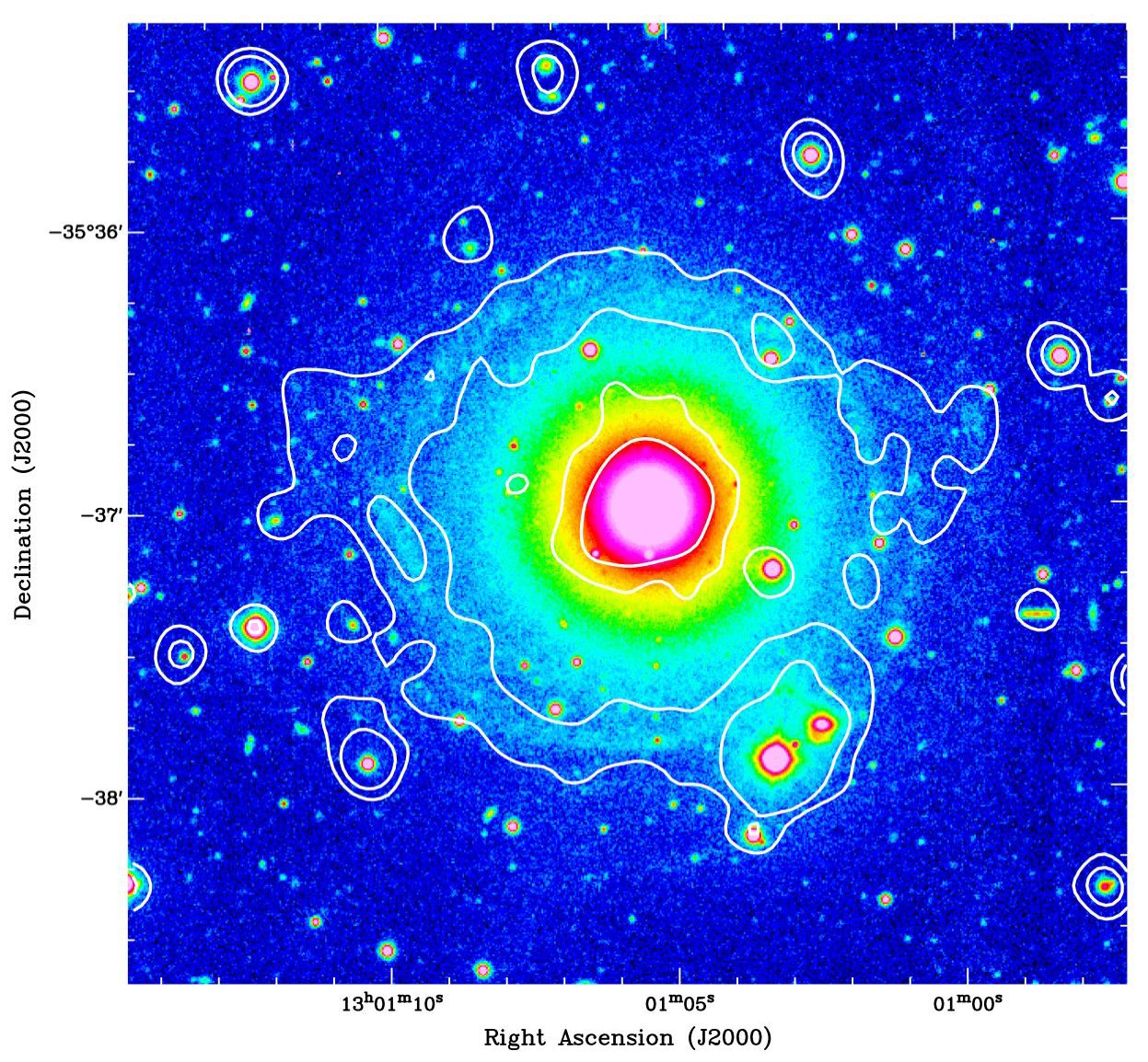} \\
\caption{$\it{Left,}$ \HI (VLA-only), $\it{center,}$ FUV, and $\it{right,}$ NUV contours overlaid on the $V$-band image of ESO 381-47. \HI contour levels are (23, 35, 46) $\times$ 10$^{19}$ cm$^{-2}$, FUV contours are (1.8, 3.6) $\times$ 10$^{-4}$ cps, and NUV contours are (1.2, 2.4) $\times$ 10$^{-3}$ cps. \label{3in1}}
\end{figure}

Overall, the ring appears to be more asymmetric in ultraviolet than in optical emission, though the seemingly elongated emission in both NUV and FUV in the east-west direction does correspond to diffuse starlight seen in the optical images. From Figure~\ref{3in1}, it is clear that the optical arms and ultraviolet ring are coincident. 

\begin{figure}
\begin{center}
\includegraphics[width=4.5in]{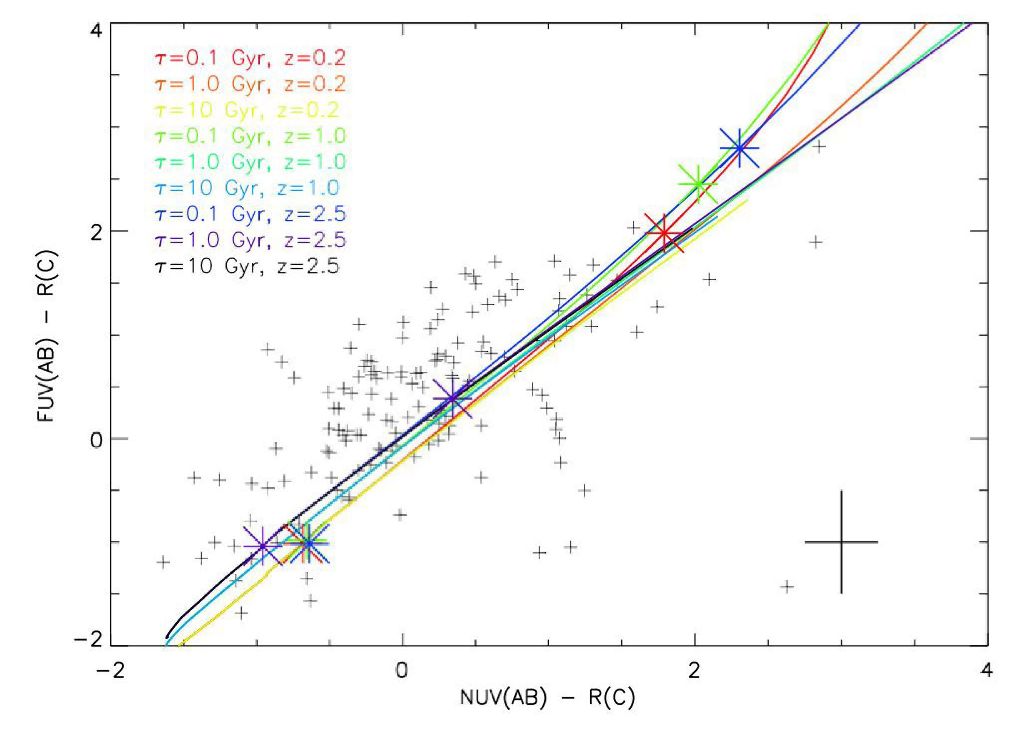} \\
\includegraphics[width=3.2in]{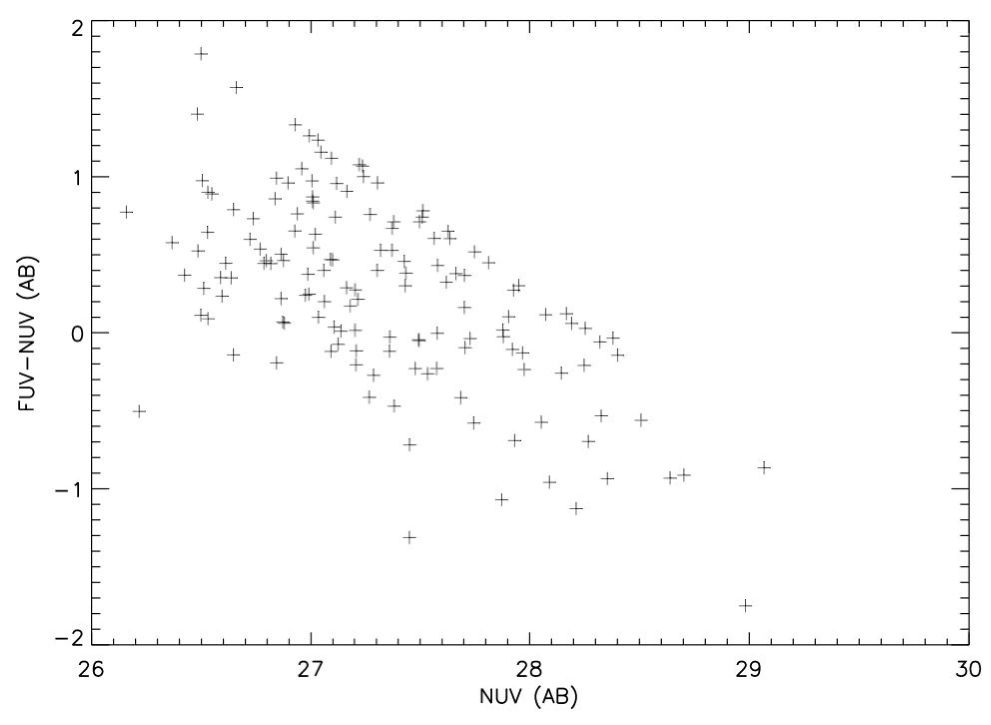}
\includegraphics[width=3.2in]{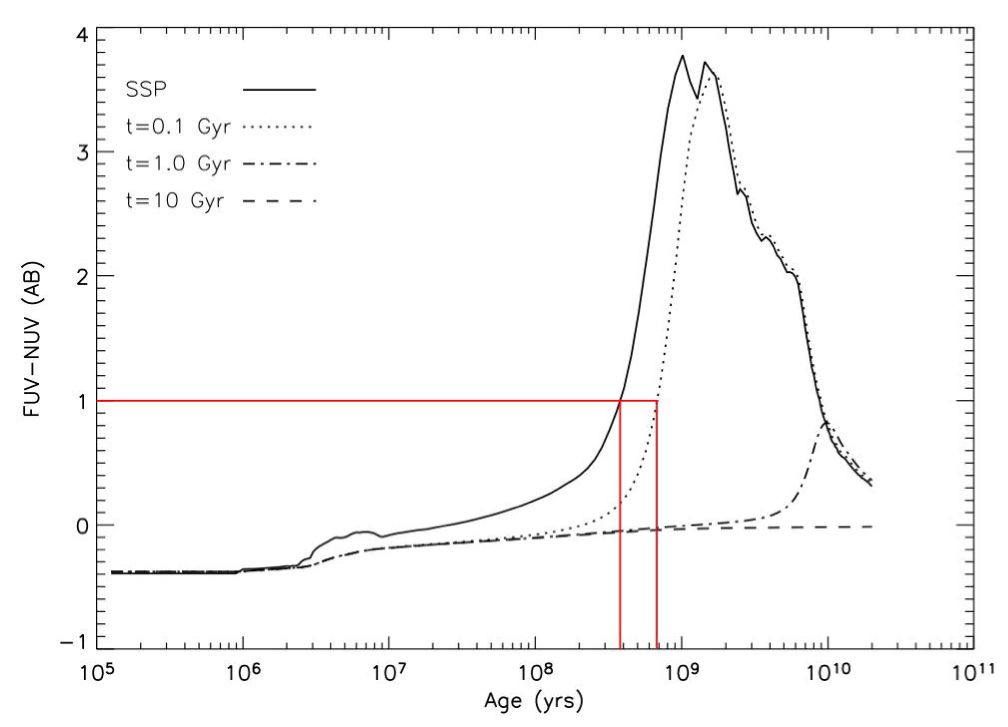}
\caption{$\it{Top:}$ $\mathrm{NUV}-R$ vs. $\mathrm{FUV}-R$ for emission in the ring. Black crosses indicate average colors within each 6$\as$ square area between the two polygons plotted in Figure~\ref{ring}. 
Colored lines indicate model ages from nine BC03 models with a burst and exponentially declining star formation with timescales of 0.1, 1.0, and 10 Gyr and metallicities of 0.2, 1.0, and 2.5 solar (\Zsun). 
Ages increase from the lower left to upper right in each model, and the asterisks in each color indicate the position where the model mean stellar age is $10^7$ and 5 $\times 10^8$ years. The uncertainties in the NUV and FUV magnitudes, which dominate the plotted colors, are shown as error bars at the lower right of the plot. $\it{Bottom~left:}$ $\mathrm{FUV-NUV}$ colors as a function of $\mathrm{NUV}$ magnitude, showing the effect of the 2$\sigma$ $\mathrm{FUV}$ cut. $\it{Bottom~right:}$ Possible star formation histories, shown at solar metallicity from the instantaneous burst and burst + exponentially declining star formation models by \citet{BC03}. The colors in the top plot help to constrain the time since the most recent burst of star formation in this panel; $\mathrm{FUV-NUV}=1$ and its corresponding ages according to the instantaneous burst and exponentially declining model with $\tau\sim 0.1$~Gyr are highlighted. \label{colorage}}
\end{center}
\end{figure}

We calculate UV-optical colors $\mathrm{NUV}-R$ and $\mathrm{FUV}-R$ for ESO 381-47 and its ultraviolet ring between the polygons shown in Figure~\ref{ring}. The emission from the ring in color-color space is shown in Figure~\ref{colorage} overlaid with nine $\mathrm{NUV}-R$ and $\mathrm{FUV}-R$ color tracks (with NUV and FUV in $AB$ magnitudes and $R_C$ in Vega magnitudes. We preserve the Vega system for our $R$ magnitudes to minimize errors in converting from one system to another; note that according to \citet{Blanton07}, R$_{AB}$ is about 0.2 magnitudes fainter than R$_{C}$ for the Sun.) These tracks are \citet{BC03} stellar population models calculated for various metallicities. Each cross represents the average color within a 6$\as$ square region -- roughly a point spread function of the ultraviolet images -- between the two polygons plotted in Figure~\ref{ring} where the average $\mathrm{FUV}$ magnitude is $<$ 28.4. Each colored track represents the color predictions of the \citet{BC03} models for stellar ages from $10^5$ to $10^{10}$ years at one of three metallicities: 0.2 \Zsun, 1.0 \Zsun, and 2.5 \Zsun. The models include a burst and exponentially declining star formation with one of three different timescales -- 0.1, 1.0, and 10 Gyr -- as shown in the plot. Ages increase from the lower left to upper right, and the stars indicate the positions where the mean stellar age in each model is 10$^{7}$  and 5$\times$10$^{8}$ years. We plot these age markers only for the three models with $\tau$=0.1 Gyr and one additional model ($\tau$=1.0 Gyr, z=2.5 \Zsun), as the age markers for the six models with 1.0 and 10 Gyr timescales are all very similar. 

While it is impossible to distinguish the best fitting model from this overlay on the data, each model is consistent with the most recent star formation occurring less than 10$^{9}$ years ago. There is an age--metallicity degeneracy such that the ages depend slightly on the assumed metallicity, but the mean formation timescale of the stars is between a few $\times 10^6$ and roughly 5 $\times 10^8$ years for the $\tau$=0.1 Gyr models, a few $\times 10^9$ years for the $\tau$=1.0 Gyr models, and 2 $\times$ 10$^{10}$ years for the $\tau$=10 Gyr models. The colors in the ultraviolet ring are on the whole redder in $\mathrm{FUV}-R$ than the models predict, which indicates that a mix of stellar populations with differing ages (older than the recent burst) may be present. 

Additionally, we compare the $\mathrm{FUV-NUV}$ colors in the ring, shown as a function of $\mathrm{NUV}$ magnitude in the second panel of Figure~\ref{colorage}, to the same \citet{BC03} models above as well as to a model of an instantaneous burst [also from \citet{BC03}] in the third panel of Figure~\ref{colorage}. Each model has solar metallicity. With a color measurement of $\sim$1.0 for the reddest stars in $\mathrm{FUV-NUV}$, we can constrain the age of the oldest stars in this most recent burst of star formation to be 350 - 700 Myr using the SSP and $\tau$=0.1 Gyr curves. The curves corresponding to more extended bouts of star formation do not produce red enough colors to match the data.

\subsection{Stellar and ionized gas kinematics}

The kinematics of the stars and ionized gas in ESO 381-47 within R$_e$/2 were presented in \citet{SerraThesis}. The stars exhibit a very small rotation of only 15 \kms, but this may be attributed to the fact that the stellar disk is nearly face on. From a detailed analysis of the stellar velocity distribution, \citet{SerraThesis} found 
further evidence for the presence of a rotating disk component coincident with a non-rotating bulge: the stellar velocity anti-correlates with the skewness of the stellar line-of-sight velocity distribution (LOSVD) as a function of radius. Given the low $\it{v/\sigma}$ ratio of early type galaxies, this behavior can only be explained with a rotating system superimposed upon a non-rotating one \citep{Bender90}. See \citet{SerraThesis} for more details. 

Metallicity measurements of ESO 381-47 were presented in S08 at the very center (within R$_e$/16) to be $\mathrm{[Z/H]}=0.18$ and within R$_e$/2 to be $\mathrm{[Z/H]}=-0.02$. The ionized gas is very faint for this galaxy and as a result was only detected within the central 10$\as$ of the galaxy by S08; no attempt was made to observe ionized gas coincident with the faint optical ring. 
In the inner regions, the kinematics of the ionized gas are decoupled from the kinematics of the stars, and instead seem consistent with an inward extrapolation of the \HI kinematics at larger radius. This will be addressed further in the following subsection (\S 4.3).

\subsection{\HI}

The integrated \HI emission of the entire field using the combined VLA+ATCA cube is shown in Figure~\ref{uvHI} overlaid on GALEX data. 
Though the companions appear to be mostly unresolved in the combined \HI data, the \HI-richness of the environment -- which will be discussed in more detail below -- is obvious. \HI mass, v$_{sys}$, and $\Delta$v for each galaxy in the field are listed in Table~\ref{opthiprops}. For simplicity, we assume in the mass calculation that all of the galaxies lie at 61.2 Mpc, the same distance as ESO 381-47.

The integrated \HI emission of ESO 381-47 from the combined VLA+ATCA cube is shown in Figure~\ref{dss} overlaid on our $V$-band and GALEX images. The diameter of the gas ring is $\sim$95 kpc in the north-south direction at its lowest contour level and $\sim$87~kpc in the east-west direction. In this figure, the VLA-only data is also shown overlaid on our $V$-band and GALEX images. 
In both maps, the gas does not appear to be totally regularly distributed, as protuberances are present on the southwestern, southeastern, and especially the northern edges of the \HI ring in the combined data and on the eastern and western sides of the high resolution data. 

\begin{figure}
\hspace{-2cm}
\includegraphics[width=7.5in]{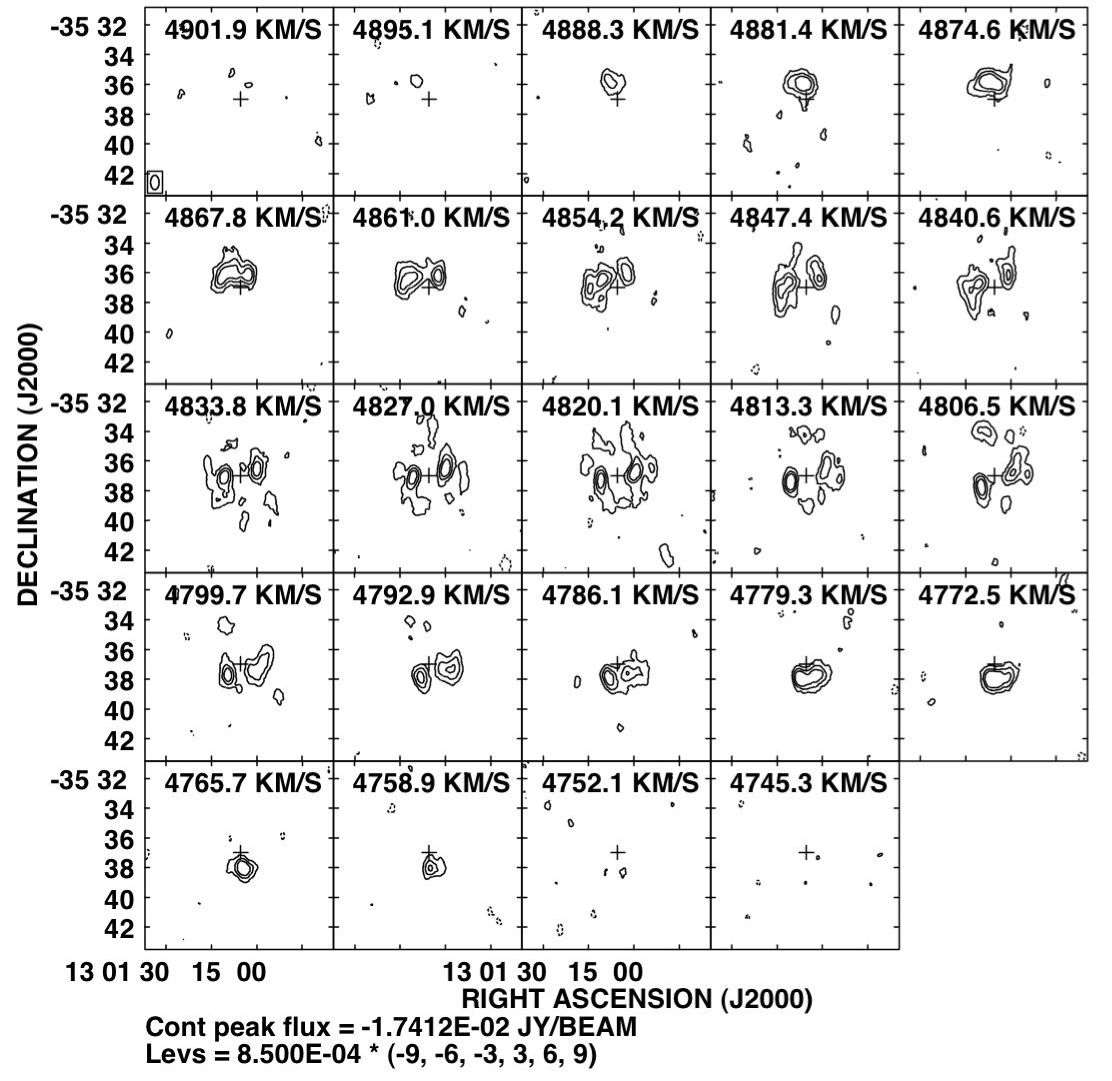}
\caption{\HI channel map of ESO 381-47. Contours are drawn at (-9, -6, -3, 3, 6, 9) $\times$ 0.85 mJy beam$^{-1}$. Positive contours are solid lines, and negative contours are dashed. Heliocentric radial velocities are given at the top of each panel. The cross in each panel represents the pointing center. \label{HIchans}}
\end{figure}

The channel maps of the VLA+ATCA \HI data are shown in Figure~\ref{HIchans}. At first glance, it appears to be consistent with a rotating ring, but upon closer inspection, emission from the entire ring appears at the systemic velocity (the 4820 \kms channel). This hints at deviations from coplanar circular gas orbits. A similar indication is provided by the presence of a large-radius NorthWestern Cloud (NWC) counter-rotating with respect to the inner \HI gas (velocity range $4793-4813$ \kms). 

\begin{figure}
\includegraphics[width=3in]{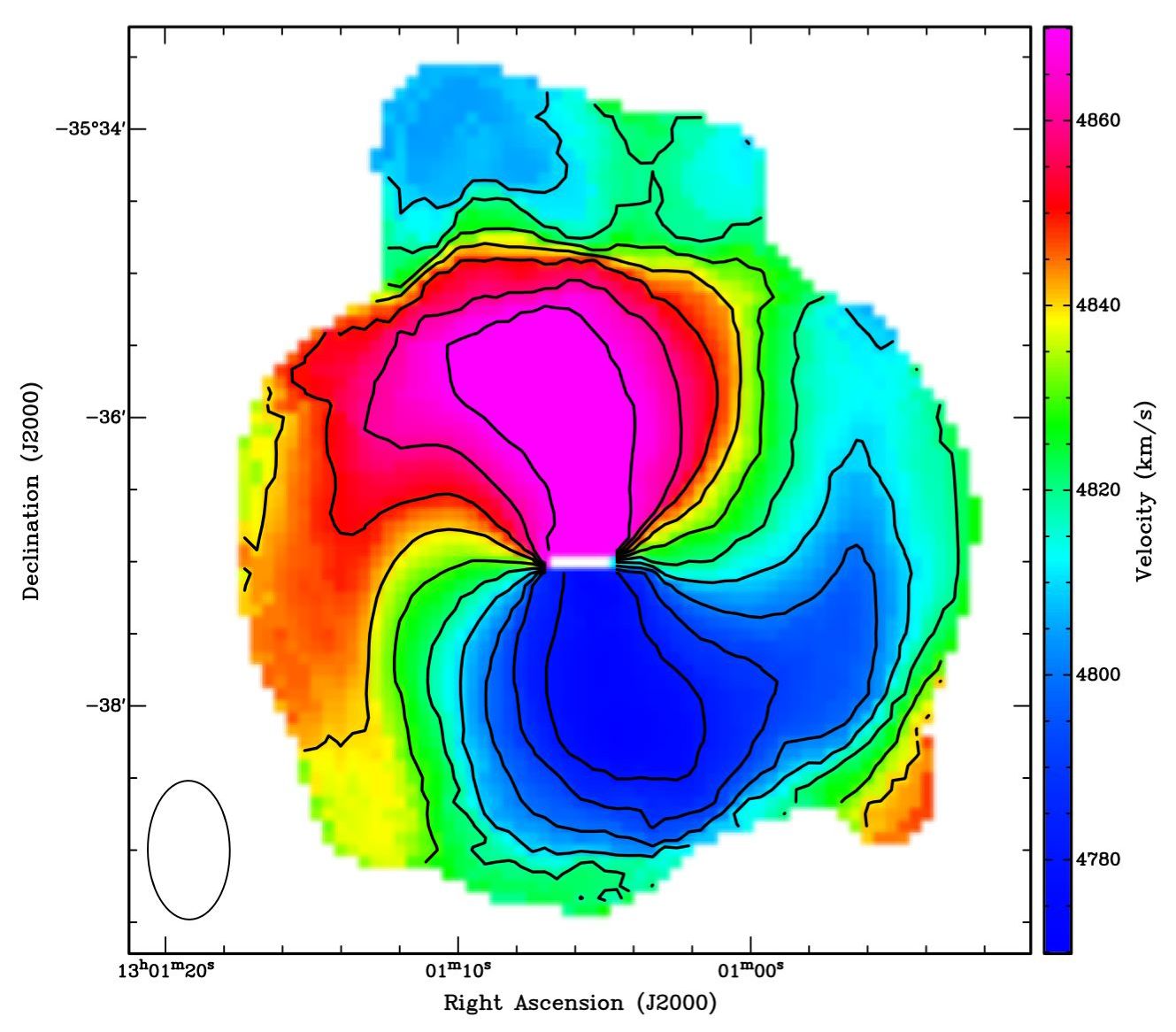}
\includegraphics[width=3in]{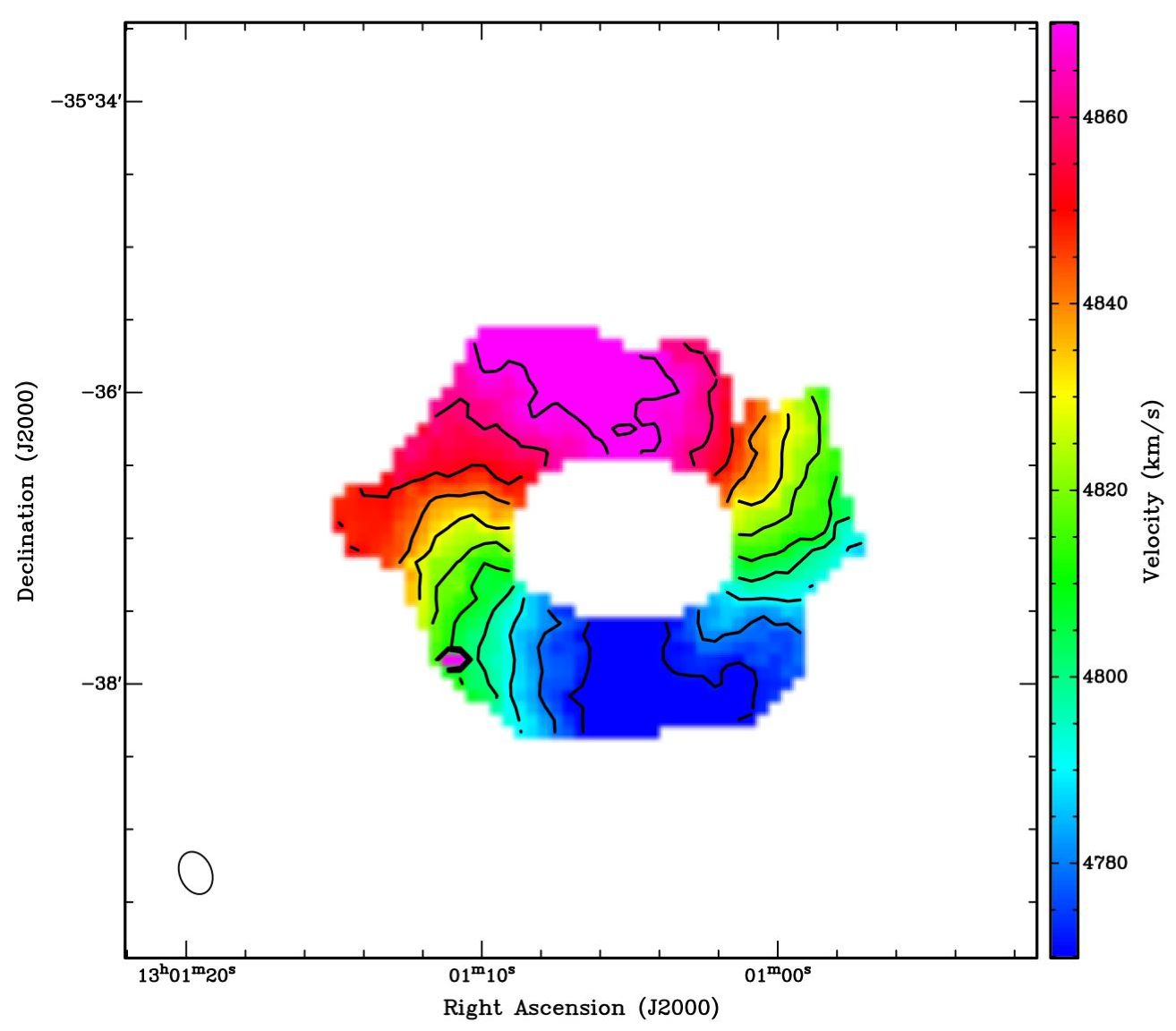}
\caption{$\it{Left,}$ intensity weighted mean velocity field of the combined VLA+ATCA data with a resolution of 34.10$\as$ $\times$ 57.71$\as$. $\it{Right,}$ same, for VLA-only data with a resolution of 13.19$\as$ $\times$ 18.15$\as$. The velocity ``wiggles" in the VLA-only data in the northeast and southwest parts of the ring are indicative of streaming motions, which we interpret as the presence of a density wave. Contours are drawn in 10 \kms increments from 4780 to 4870 \kms for the VLA+ATCA data and from 4770 to 4870 \kms for the VLA-only data. The beams are plotted at the lower left in each panel for reference. \label{velfields}}
\end{figure}

We show the velocity field of the VLA+ATCA \HI data in ESO 381-47 in the left panel of Figure~\ref{velfields}. The large scale behavior of the gas is evident in the VLA+ATCA velocity field, which exhibits the ``S" shape commonly associated with warped disks. The blueshifted arm can be seen to terminate in the NWC on the opposite side of the galaxy, assuming that some of the intervening emission is too faint or clumpy to be detected, and the redshifted arm also exhibits a redshifted cloud opposite its position suggesting that faint connecting emission is also missing from this arm. 
At the outer edge of the VLA+ATCA data, the velocity of the gas can be seen to return to systemic, or close to it, suggesting that the disk at that location returns to being close to face-on from being inclined to the line of sight. The blue and red tails at the end of the arms, discussed above, overshoot the systemic velocity and continue to warp in their respective directions.

The channel maps and velocity field of the combined \HI data indicate that the gas is not simply in coplanar circular motion, and we present a warped disk as a possible explanation. To test this hypothesis, we build a simple tilted-ring model cube \citep{Rogstad74} that captures the basic kinematical and morphological properties of the combined \HI data. To do so we use the combined VLA+ATCA data cube as a reference, since it contains important information on the large-radius, faint emission. In building the model, we assume that the HI is in equilibrium with the potential and moves along concentric circular orbits centered at the optical center of the galaxy. The plane of the circular orbits is allowed to vary with radius (giving the so called tilted rings). We assume that the gas velocity dispersion is 10 \kms. Changing this value slightly does not significantly affect the result of the modeling. Given the nearly-zero inclination of the HI system, accurately determining the gas rotational velocity and its variation with radius is not possible (i.e., error bars would be very large). Therefore, we assume a flat rotation curve.

\begin{figure}
\includegraphics[width=6in]{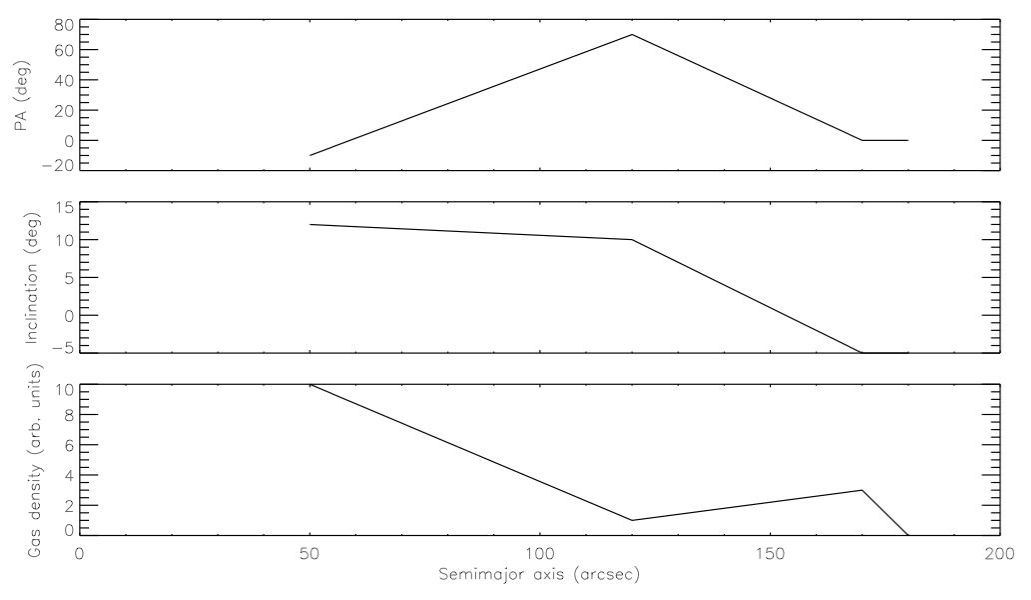}
\caption{\HI model parameters position angle, inclination angle, and gas density as a function of semi-major axis. \label{3panels}}
\end{figure}

\begin{figure}
\includegraphics[width=7in]{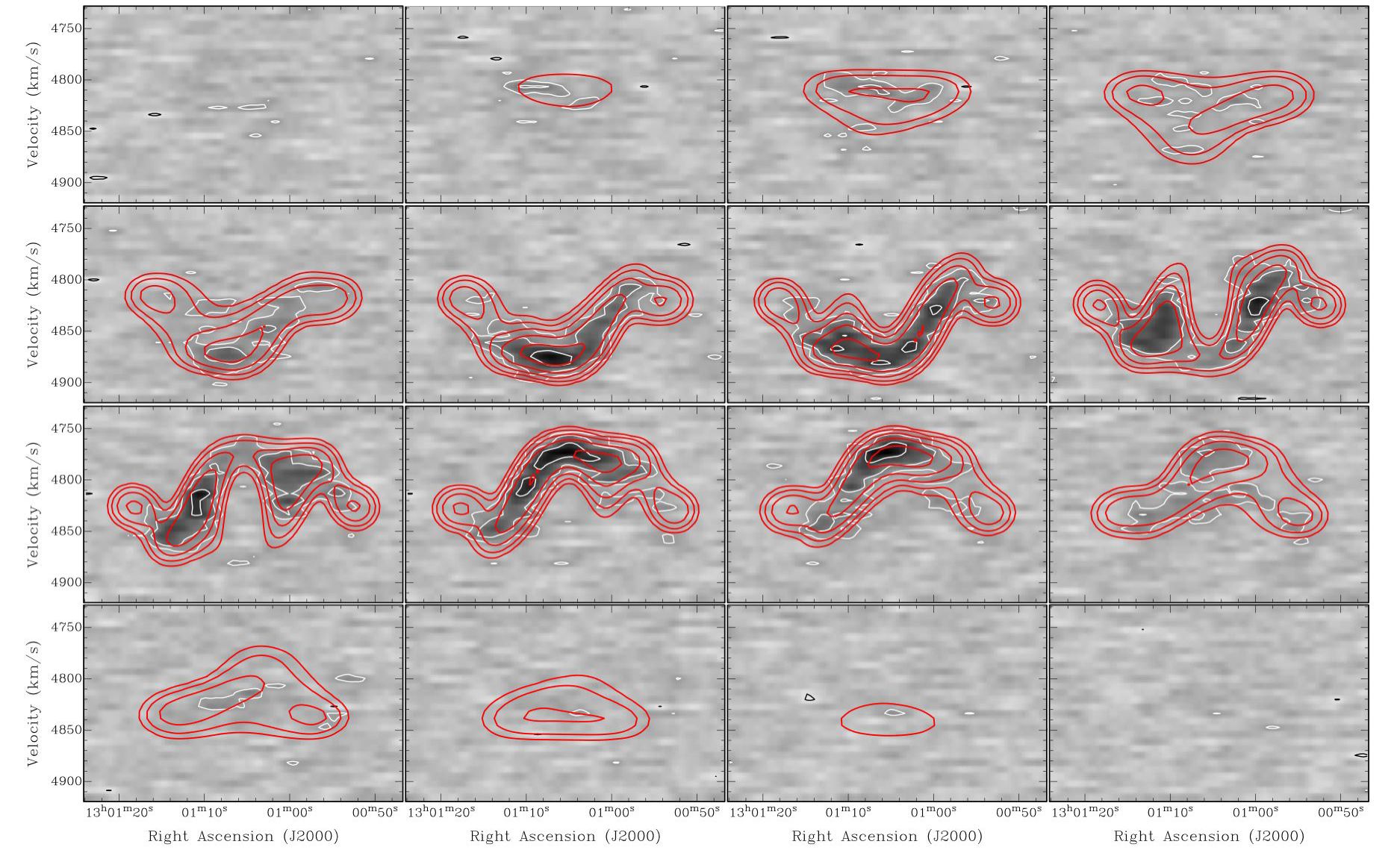}
\caption{\HI model, shown by the red contours, overlaid on the combined \HI data, shown by the gray scale, white (positive) contours, and black (negative) contours. The simple model does a good job except at large radii where the gas is very faint and asymmetric. Each panel is a position-velocity diagram drawn at fixed declination along PA=90$\dg$. The declination decreases from left to right, top to bottom in steps of 30$\as$ (i.e., the top-left diagram is at the northern edge of the \HI distribution, and the bottom-right diagram is at the southern edge). The first panel (top-left) is drawn at Dec=-35$\dg$33$\am$45$\as$. \label{modeloverlay}}
\end{figure}

Although the data cube shows a fairly asymmetric (particularly in the outer regions) and clumpy HI distribution, we can satisfactorily model it with a symmetric, smooth model. We find that a warped ring where position angle, inclination, and gas density vary as in Figure~\ref{3panels}, and with a rotational velocity of 260 km/s, is a good match to the data. We point out that this is not a best fit, but rather a simple model that, with a few parameters, reproduces the observations well. Figure~\ref{modeloverlay} shows a comparison of the data (grayscale and white contours) to the model (red contours). The differences between model and data at large radius are due to our attempt to model the emission in these regions also, where the gas is very faint and asymmetric. 

We tested our simple model together with Gyula J{\'o}zsa using his TiRiFiC $\chi^{2}$-minimization code \citep{Jozsa07}. TiRiFiC determines local minima of the $\chi^{2}$ distribution when fitting data with tilted ring models. Our model, the parameters of which were not derived from minimization, is virtually indistinguishable from what we find when running TiRiFiC; in fact, the only difference between our simple model and the best-fitting TiRiFiC model is that ours reproduces the outer faint and asymmetric region slightly better.

Our model does not recover the \HI kinematics on smaller scales, which can be studied in the VLA-only data shown in the right panel of Figure~\ref{velfields}. Here, deviations from circular motions can be seen in the iso-velocity contour ``wiggles", especially on the northern and southern sides of the ring. This behavior is often seen near spiral arms. 
The larger beam used in the lower resolution images smooth out the wiggles, and for the model we use the low resolution data in order to recover the largest scale \HI structure.

\begin{figure}
\includegraphics[height=2in]{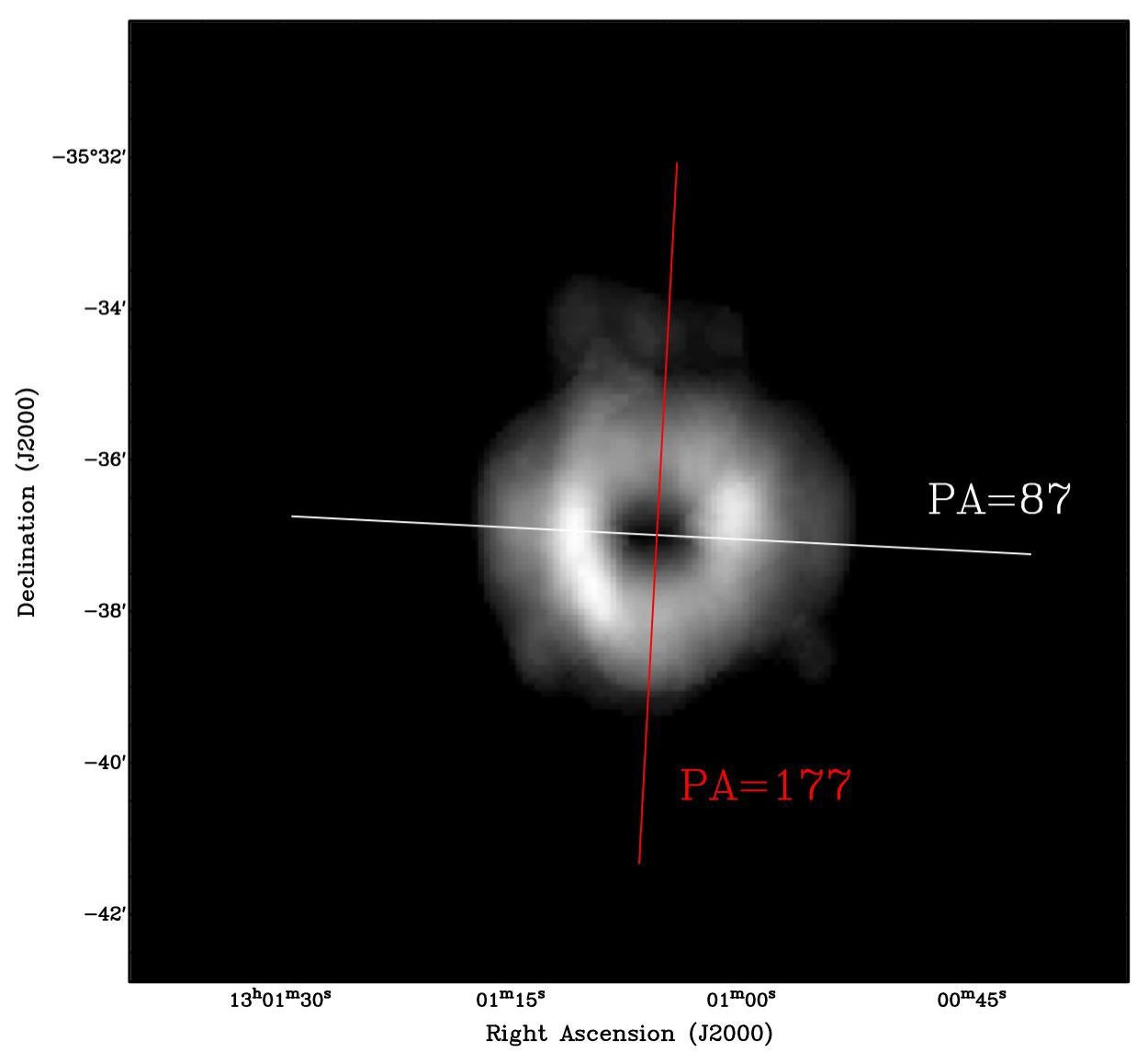}
\includegraphics[height=2in]{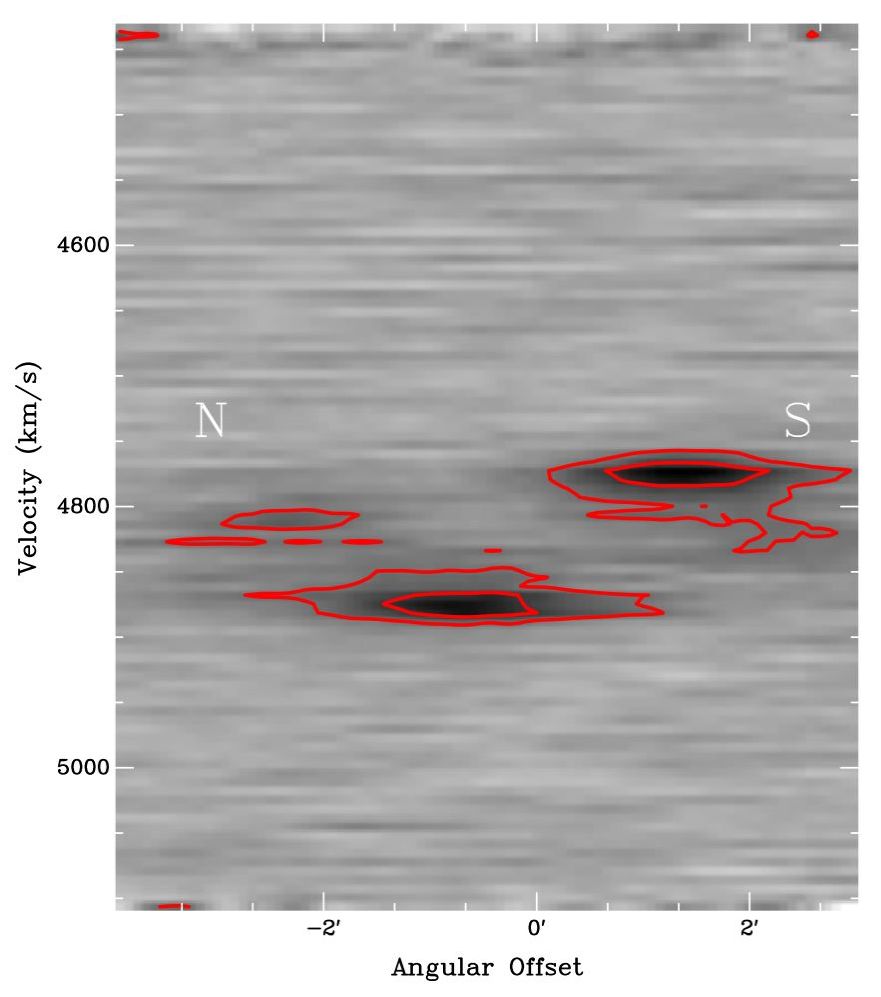}
\includegraphics[height=2in]{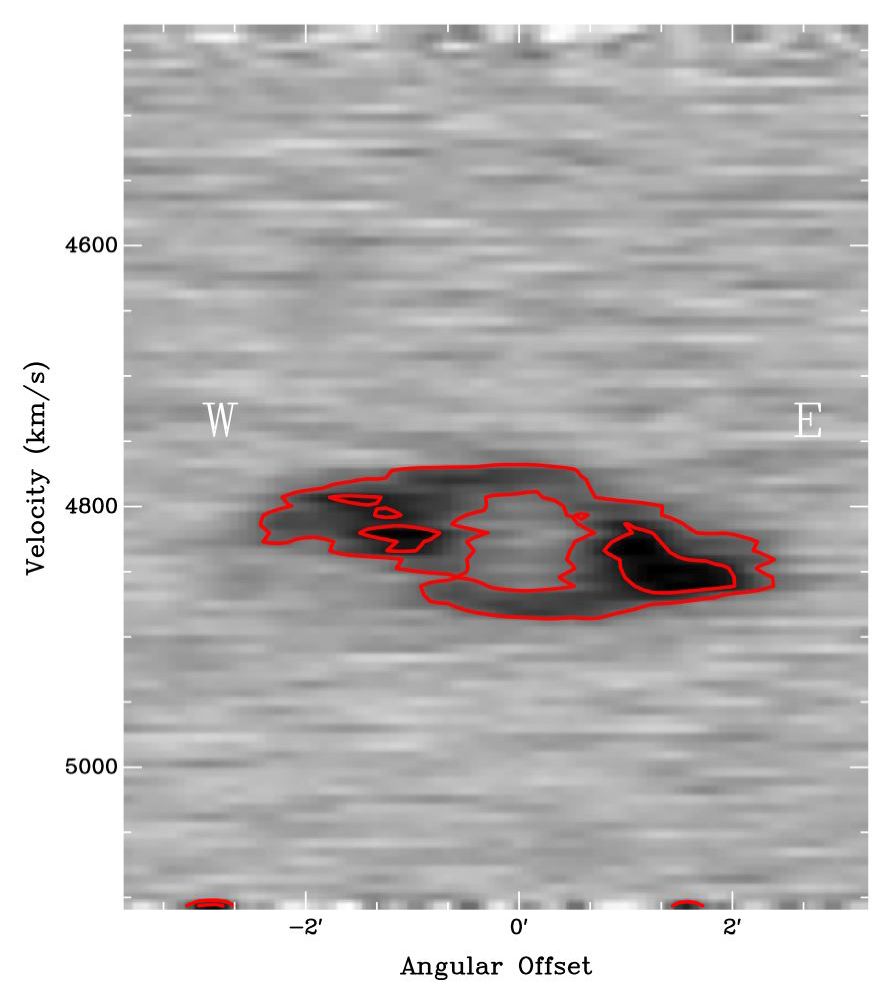} \\
\begin{center}
\includegraphics[width=6in]{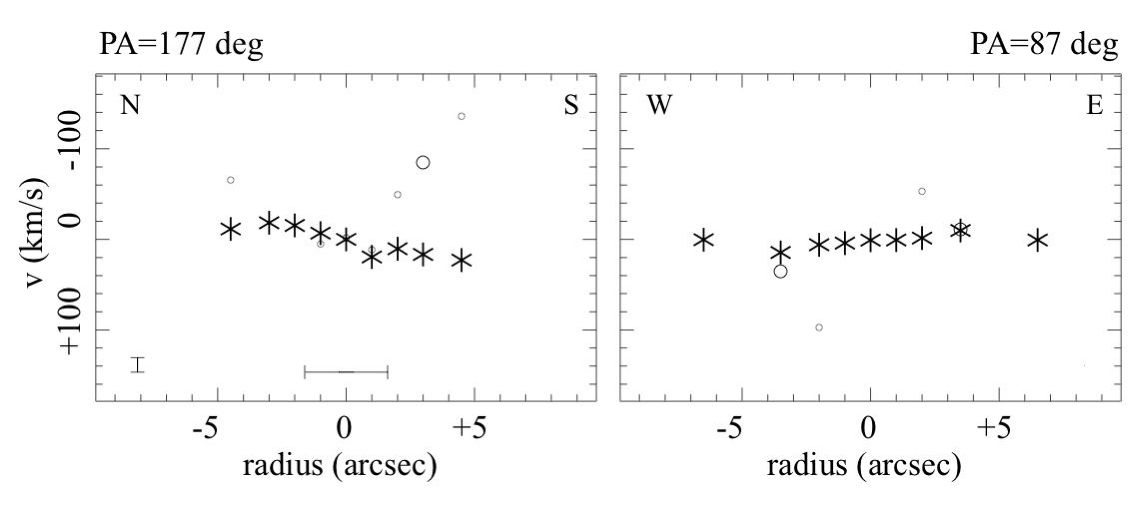} \\
\caption{Position-velocity diagrams for two slices, PA=177$\dg$ and PA=87$\dg$, through the \HI ring of ESO 381-47. $\it{Left,}$ total \HI with the PA=87$\dg$ cut overlaid in white and the PA=177$\dg$ slice shown in red. $\it{Center,}$ PV diagram of the PA=177$\dg$ cut. $\it{Right,}$ PV diagram of the PA=87$\dg$ cut. The warped nature of the disk is apparent especially in the right panel, as the ends of the arms return to the systemic velocity of the system. Contours are at 5, 10~mJy. $\it{Bottom,}$ faint [OIII] ionized gas is shown to be consistent with an inward extrapolation of the \HI. Stars indicate the stellar velocities and circles the velocity measurements of [OIII] where small circles indicate 3$<$A/N$<$5 and large circles indicate A/N$>$5; A is the amplitude of the best-fitting Gaussian profile to the detected [OIII] line and N is the noise. The zero velocity corresponds roughly to the systemic velocity of ESO 381-47. Horizontal and vertical bars in the left panel are the 1 kpc scale (at 63.7~Mpc, the distance assumed in \citet{SerraThesis} for ESO 381-47) and the typical error on the stellar velocity. North/south and east/west are labelled on each plot for ease of comparison. \label{pvs}}
\end{center}
\end{figure}

Finally, it is illustrative to investigate position-velocity (PV) diagrams to analyze the kinematics of the \HI present in ESO 381-47. Figure~\ref{pvs} shows two PV diagrams of the VLA+ATCA data along PA=177$\dg$ (top-middle panel) and PA=87$\dg$ (top-right panel). The two slices are shown overlaid on the total \HI image in the top-left panel. Here again the presence of the warp is displayed in the counter rotation of the extended arms to each side of the galaxy, particularly in the PA=87$\dg$ PV diagram. At the end of each arm, the velocity curls back to the systemic velocity of the system, as is seen in the velocity fields in Figure~\ref{velfields}. 

For comparison, Figure~\ref{pvs} also shows the ionized gas within the central 10$\as$ of ESO 381-47, taken at the same position angles as the PV cuts above \citep{SerraThesis}. Stars indicate the stellar velocities and circles the velocity measurements of [OIII] where small circles indicate 3$<$A/N$<$5 and large circles indicate A/N$>$5; A is the amplitude of the best-fitting Gaussian profile to the detected [OIII] line and N is the noise. The zero velocity corresponds roughly to the systemic velocity of ESO 381-47. It appears that the ionized gas is decoupled from the stellar kinematics and, while very much interior to the \HIcom it is consistent with an inward extrapolation of the \HI kinematics at larger radius. In fact, the consistency between the kinematics of outer \HI and inner ionized gas motion has been known for a long time (\citealt{vanGorkom97} and references therein) and was a general conclusion in \citet{SerraThesis} and \citet{Morganti06} based on large samples of early type galaxies. In the PA=177$\dg$ case, it seems that the ionized gas is generally approaching in the north and south of the system, which only matches the southern \HI behavior. At PA=87$\dg$, the ionized gas appears to be approaching to the east while receding to the west, and in both directions the gas returns to the systemic velocity at its largest detected radius; the \HI at much larger radius decreases to systemic on the east and increases to systemic on the west, in a sense continuing the trend of the ionized gas. However, a definitive conclusion is impossible as the physical separation between the ionized gas and \HI ring is at least 10~kpc. Neither the ionized gas nor the \HI exhibits the same kinematics as the stars.

\begin{figure}
\includegraphics[width=2.5in]{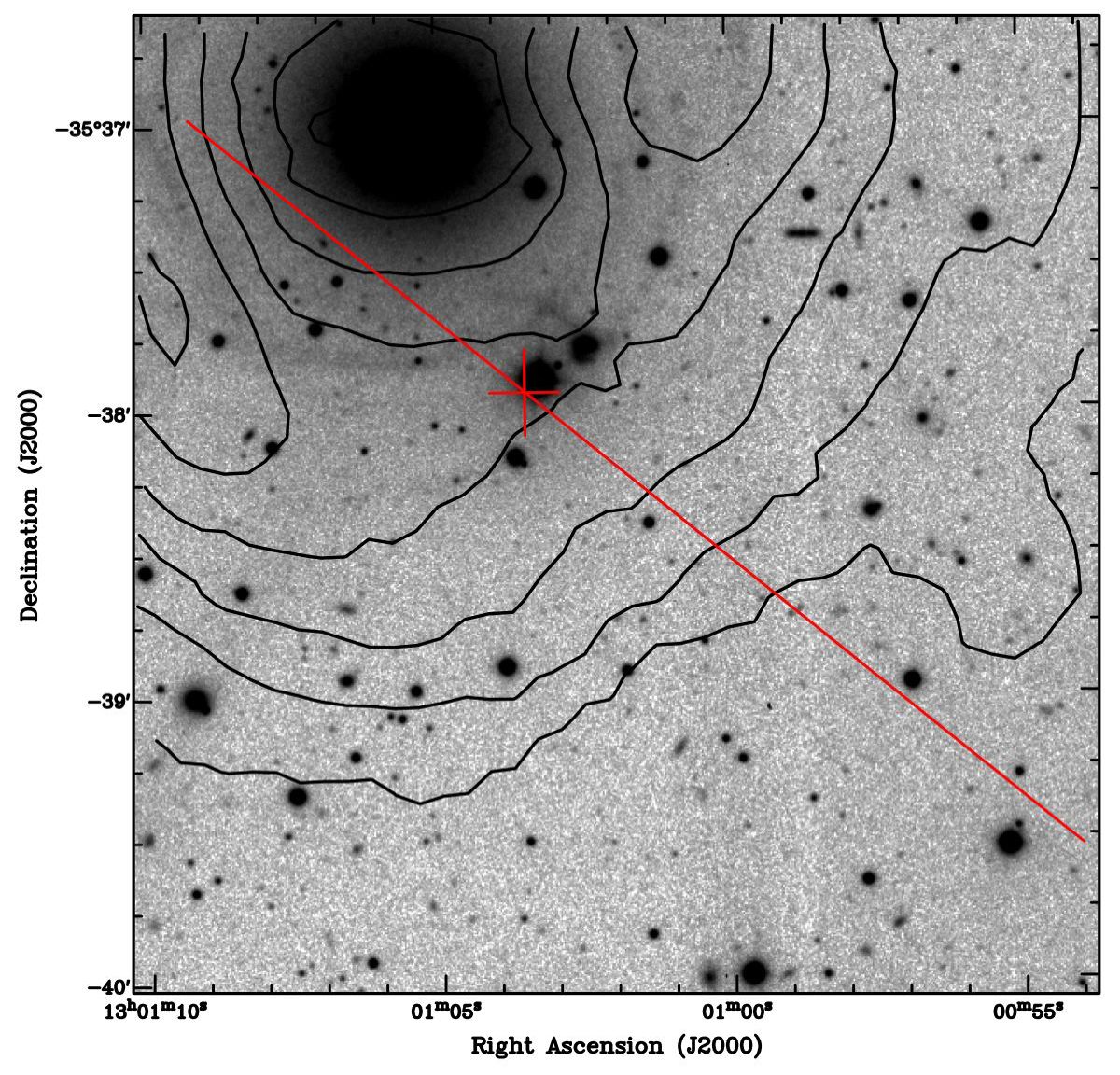} 
\includegraphics[width=2.5in]{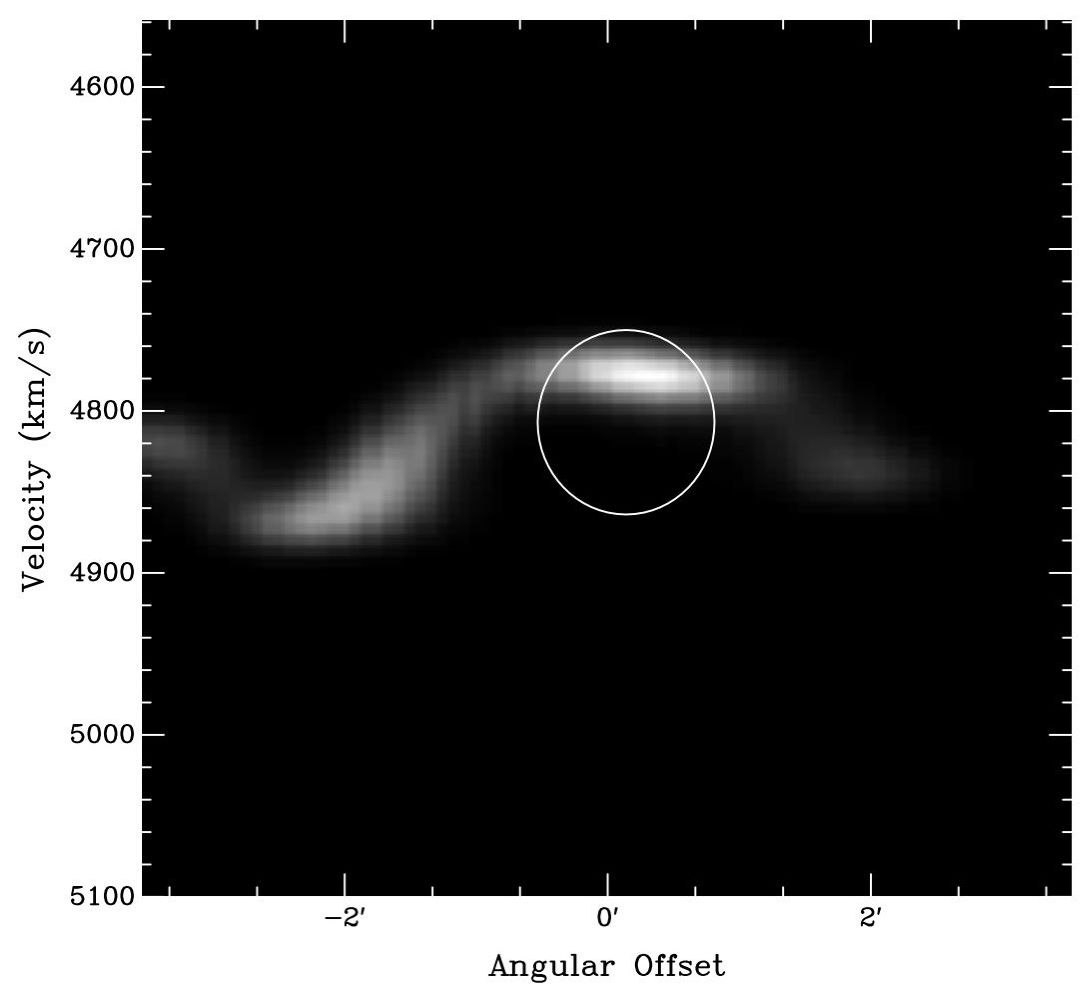}\\
\includegraphics[width=2.5in]{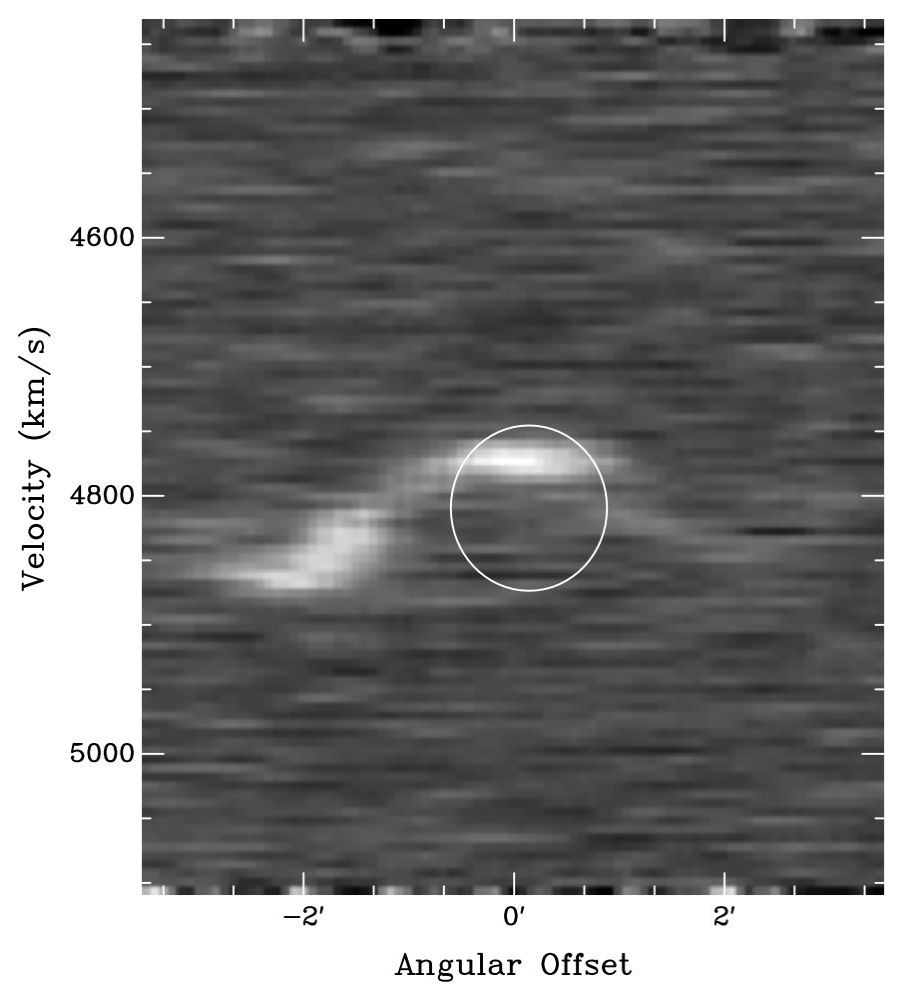}
\includegraphics[width=2.5in]{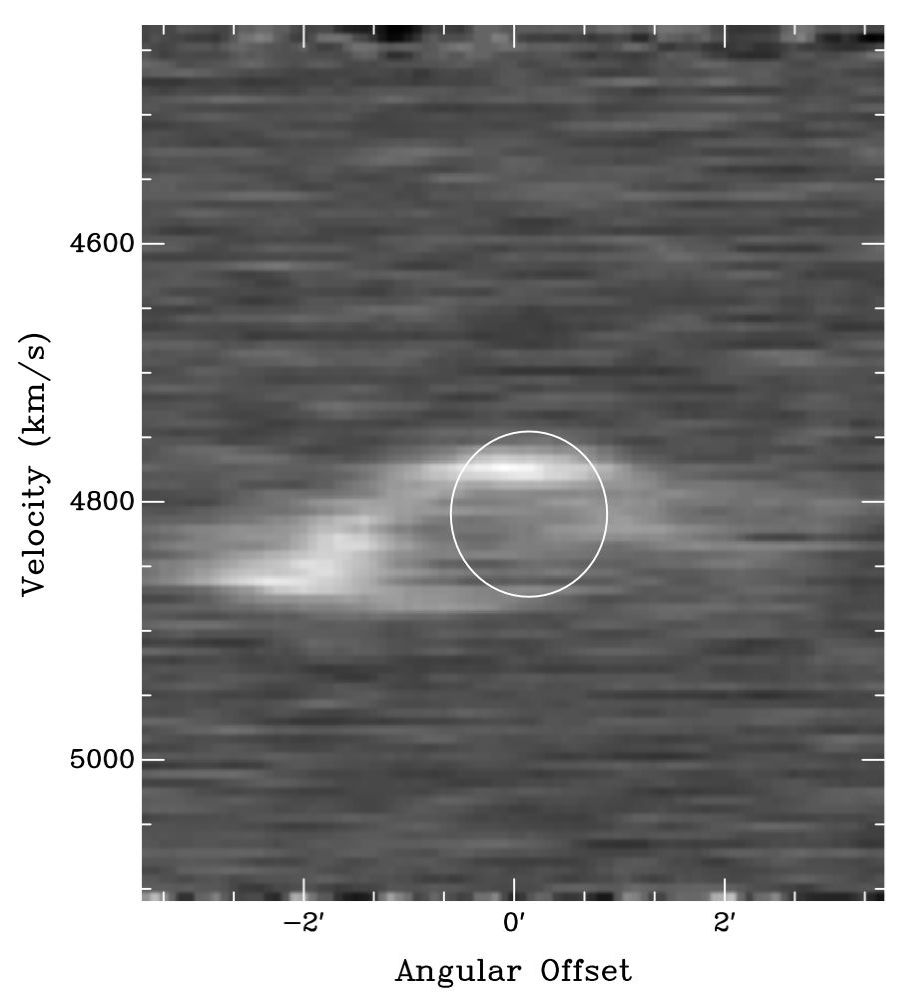}
\caption{Position-velocity diagrams for the red slice through the \HI ring of ESO 381-47 (top left). In the bottom panels, the joining of two separate structures can be seen corresponding roughly to the optical position of ESO 381-47A in a high resolution VLA+ATCA cube (bottom left) and a low resolution VLA+ATCA cube (bottom right). White circles highlight roughly the position of the optical companion and the possible weak discontinuity between the two structures; the center of these circles is the same as the position of the red cross in the image. In the top right panel, the model cube with the same PV cut does not recover the faint emission reaching toward the center of the circle. \label{pvs2}}
\end{figure}

In Figure~\ref{pvs2}, the \HI in black contours is shown with a different red slice; the cut is overlaid on the V-band data in this case in order to see its position relative to ESO 381-47A. The slice falls just south of this small galaxy in order to display an interesting feature in the cold gas. Two cubes are presented in the bottom panels: a high resolution PV diagram and low resolution PV diagram, both made from the VLA+ATCA data. The position of the red cross in the image (just south of the optical companion) is shown in each PV diagram as the center of the white circles. In the bottom panels of Figure~\ref{pvs2}, this position is nearly coincident with the apparent juxtaposition of two potentially separate gas structures. The first structure is that of the ring in two bright sections centered at offset=0 and offset=-2$\am$ (the same as those seen in Figure~\ref{pvs}), but there appears to be a second, more diffuse tail-like structure stretching to positive offset. In the high resolution PV diagram, the shape of the tail as a separate structure is easier to identify. In both \HI data PV diagrams, this more diffuse tail reaches the position of the optical companion at the center of the white circle, but with the same slice through the model shown in the top right panel, the shape is different -- the faint tail is still present, but it does not reach the center of the circle. If this feature is real, which is difficult to determine due to the faint nature of the emission, then it would imply that the cause is separate from the warped ring. 

\subsection{Correlating UV and \HI}

\begin{figure}
\includegraphics[width=5in]{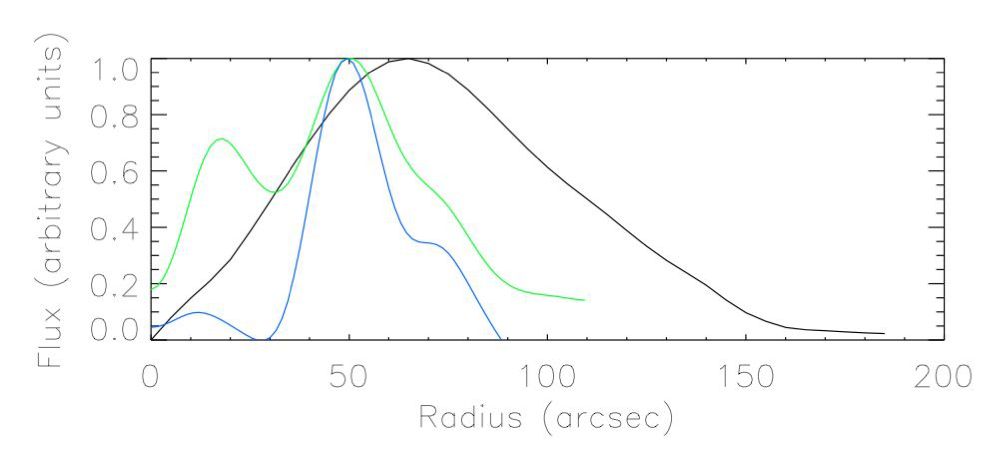}
\caption{Azimuthally averaged radial profiles of NUV (green), FUV (blue), and HI (black). We show the VLA-only data in this HI profile. The NUV and FUV profiles have been smoothed to a resolution of 15$\as$ in order to be directly comparable to the resolution of the VLA (\HI) data, which has a resolution element of 13.19$\as$ $\times$ 18.15$\as$. \label{radprofile}}
\end{figure}

In the ring, the peak of the \HI emission lies at slightly larger radii than the ultraviolet emission except for one coincident clump of UV and \HI in the northwestern corner of the ring. The azimuthally averaged radial profiles of the VLA-only HI emission and FUV/NUV profiles smoothed to 15$\as$ resolution (so as to be comparable with the \HI) plotted in Figure~\ref{radprofile}, clarify this point. The peak of the \HI surface density lies at 65$\as$ while the peaks in the other three bands lie around 50$\as$. Also, in Figure~\ref{3in1}, the highest surface density \HI contours and ultraviolet contours are overlaid on the optical image of ESO 381-47 from Figure~\ref{vband}. The FUV and NUV contours are coincident with the outer ring of stars seen in the $V$-band, while the \HI contours reach their peak just outside of the stellar ring. 

A configuration of \HI where the peak of the intrinsic surface density of gas is located precisely at the inner edge of the gas ring but appears to be at a larger radius due to the smearing effect of the beam is possible. However, the resolution of the \HI and UV data shown in Figure~\ref{radprofile} is directly comparable, making the difference in the location of the peaks significant.
Another explanation for the mismatch of the UV and \HI peaks is that the peaks actually do occur at the same location, but dust is obscuring the optical and UV light at the location of the densest neutral gas. However, the non-detection of the ring around ESO 381-47 in the IR broadband surveys would exclude obscuring hot dust. We also see no evidence for dust in the UV-optical color image, though the signal-to-noise is degraded in this image at the position of the ring as a result of its optical faintness. A real offset between the peaks in UV and \HI supports a scenario in which the motion of a density wave through the region acted as a trigger for the star formation; this will be discussed further in \S 5.3.

\subsection{Environment}

\begin{figure}
\includegraphics[width=5in]{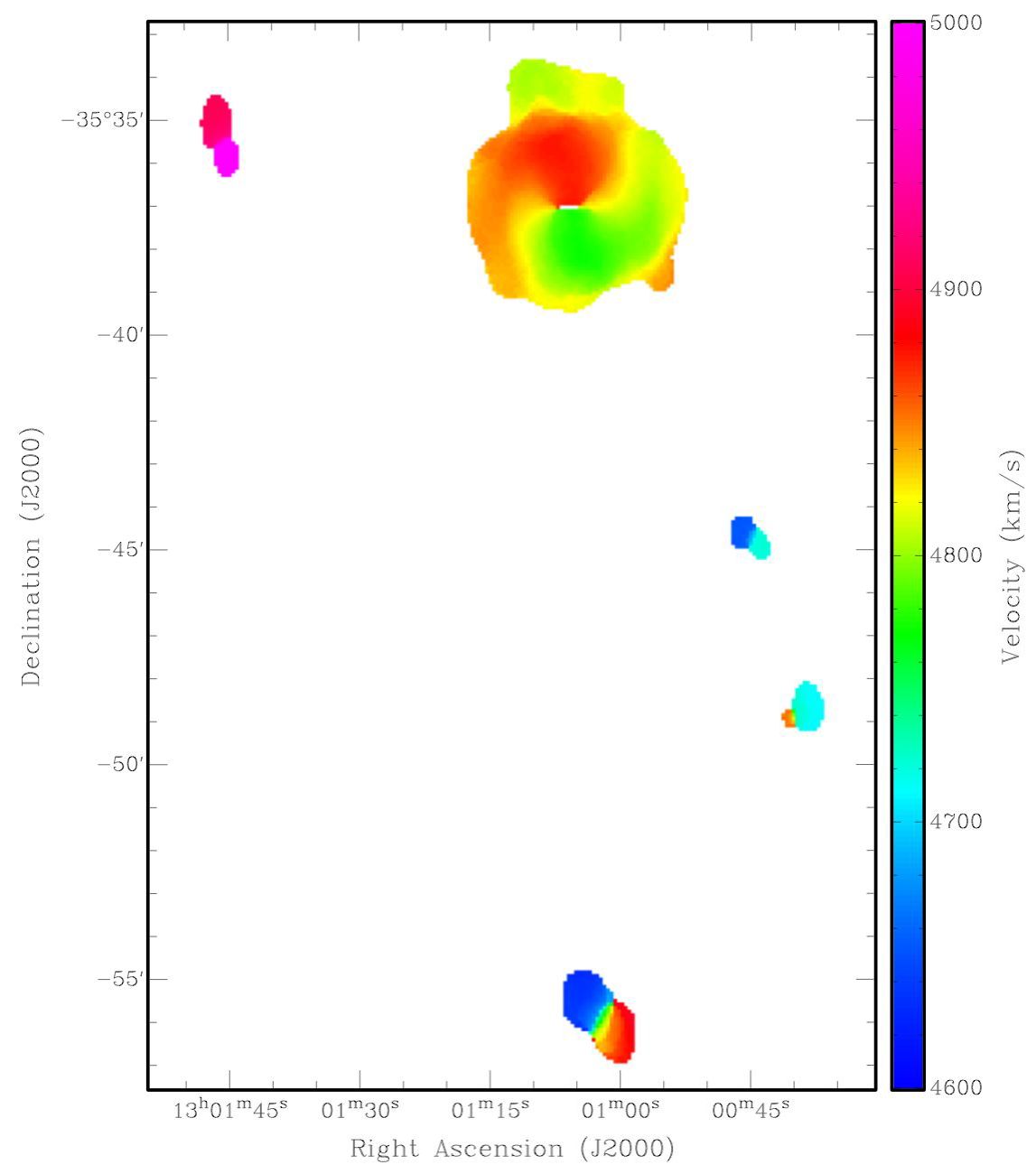}
\caption{\HI intensity weighted mean velocity field of entire group, showing the continuity in velocity between the five galaxies. \label{velgroup}}
\end{figure}

The environment around ESO 381-47 is rich with \HI, as shown in Figure~\ref{uvHI}. The four galaxies in the ESO 381-47 group, namely 2MFGC 10338, ESO 381-43N, ESO 381-43, and ESO 381-46, are each detected in the FUV band. They appear to be normal, spiral galaxies with \HI and recent star formation. A velocity field of the group is shown in Figure~\ref{velgroup}.

From the velocity field of the group, it is clear that there exists a coherent velocity structure among the gas-rich galaxies surrounding ESO 381-47. As we move from one end of the field to the other (i.e. from 2MFGC 10338 to ESO 381-43), the gas velocities seem to move smoothly from one galaxy to the next, which can be seen particularly in the continuity of each galaxy's systemic velocity, listed in Table~\ref{opthiprops}. ESO 381-46 exhibits velocities across the entire range spanned by the other four systems in the group. 
These galaxies may indeed be members of a gravitationally bound group, even though previously only ESO 381-47 and ESO 381-46 were identified as group NOGG P2 695 by \citet{Giuricin00}. 
\citet{Ragone06} also identified a group in the field of ESO 381-47 at a velocity of 5064 \kms with four members brighter than apparent magnitude $\it{b}$=19, but they do not specify which four members they detect.

\section{Discussion}
\label{discussion}

In the preceding sections, we present evidence that ESO 381-47 is an early type galaxy surrounded by a massive, warped \HI ring and a UV-bright ring of young stars located just inside the densest region of the cold gas. Superposed on this ring is a small companion galaxy which appears to be undergoing a strong interaction and losing much of its stellar material in a tidal tail which points toward ESO 381-47 at the center of the ring. 
The ring of \HI is strongly warped but is well modeled by regular rotation except on the smallest scales (as shown in \S 4.3), and the kinematics of the ionized gas at the center of the galaxy are consistent with an inward extrapolation of the kinematics of the cold gas. 

We present new optical, UV, 1.4 GHz continuum, and HI observations in this paper in order to address three main questions. First, is it surprising that there is a ring of recently formed stars around an otherwise old, normal early-type galaxy? Second, can we relate the origin of the ring to the underlying optical stellar component, or perhaps to an event which also left its signature in the cold gas? And finally, what can we learn from the current state of the \HI around ESO 381-47? 

\subsection{Star formation and the ring}

Radio continuum measurements are a useful tracer of recent star formation, as the continuum is sensitive to relativistic particles accelerated by the supernova explosions of dying O stars. It is thought that this is the reason why such measurements correlate remarkably well with far infrared measurements of galaxies. \citet{Yun01} use the FIR-radio correlation in order to predict the amount of star formation present purely from a radio luminosity at 1.4~GHz with the equation 

\begin{equation}
SFR~(M_{\odot}~year^{-1}) = 5.9 \pm 1.8 \times 10^{-22}~L_{1.4~GHz} (W~Hz^{-1}), 
\end{equation}

\noindent where L$_{1.4~GHz}$ is the 1.4~GHz luminosity. We use the upper limit on the 1.4 GHz continuum emission from the ring around ESO 381-47 (2.3 mJy) to calculate an upper limit to the massive star formation rate present of 0.61 $\pm$ 0.19 \Msun~yr$^{-1}$. We include the estimate here for comparison with our UV-derived star formation rate (SFR). 

Using the NUV and FUV imaging, we can also put constraints on the recent star formation in the ring. 
Assuming a Salpeter IMF and a relatively flat UV spectrum over both ultraviolet bands, the relation published in \citet{Kennicutt98} is

\begin{equation}
SFR~(M_{\odot}~year^{-1}) = 1.4\times 10^{-28} L_{\it{v}}~(ergs~s^{-1}~Hz^{-1}), 
\end{equation}

\noindent where L$_{\it{v}}$ is the UV luminosity in either NUV or FUV; compared to the precision of the star formation rate derived in this way, the difference between NUV and FUV is sufficiently small. This relation applies to systems exhibiting continuous star formation for 10$^{8}$ years or longer, which is consistent with the star formation timescales that we find in the ring, as shown in Figure~\ref{colorage} and discussed in \S 4.1. The detected NUV flux integrated over the ring allows us to estimate a dust-corrected star formation rate of 0.19 $\pm$ 0.085 \Msun~yr$^{-1}$. 

The UV-calculated SFR value in the ring is entirely consistent with the upper limit set by our observations at 1.4 GHz, even though the two trace different stellar populations. Radio continuum measurements are dominated by relativistic electrons from the supernovae of massive O stars tens of Myr after a starburst, while less massive B stars live longer and can be detected by our ultraviolet imaging. 

To analyze whether the presence of the UV star formation is surprising, we calculate the threshold gas density for star formation using the relation for the radial dependence of the critical density in a disk, assuming a flat rotation curve, published in \citet{Kennicutt89}: 

\begin{equation}
\Sigma_{gas, crit}~(M_{\odot}~pc^{-2}) = 0.59~\alpha~V~(km s^{-1}) / R~(kpc),
\end{equation}

\noindent where $\alpha$ = 0.67, while V and R are the velocity and radius of the measurement. We calculate this threshold \HI surface density at three positions in the UV/\HI overlay to compare the calculated threshold with the actual measured gas density.

First, along the ultraviolet ring: the southernmost point in the UV ring is located within the 3.5 $\times$ 10$^{20}$ cm$^{-2}$ VLA contour (see Figure~\ref{dss}). This corresponds to an observed gas surface density ($\Sigma_{gas, obs}$) of 4.1 \Msun~pc$^{-2}$. Here and below, $\Sigma_{gas, obs}$ is our observed total neutral hydrogen density scaled by a factor of 1.45 to account for helium and heavier elements in the same fashion as \citet{Kennicutt89}. The molecular fraction in the \HI ring is probably low, but CO observations would be necessary to know this definitively. 
At this location, V = 260 \kms~from our model and R $\sim$ 15~kpc. This yields a $\Sigma_{gas, crit}$ of 6.9 \Msun~pc$^{-2}$, a factor of almost 2 greater than the surface density that we observe. This suggests that star formation would not be expected where we observe it, but it is possible that, due to the likely clumpy structure of the gas, the spatial resolution of the observing beam will smear out denser pockets of \HIcom which could be at or above the calculated threshold.


Second, we calculate $\Sigma_{gas, crit}$ within the highest column density gas contours. In some areas, the densest gas is coincident with UV emission (i.e. the eastern edge of the ultraviolet ring in the VLA image of Figure~\ref{dss}), but over several other areas, the densest contours overlap with the UV emission but lie mostly at larger radius (i.e. the southwestern, southeastern, and northern edges of the ring). For instance, we examine the southeastern edge of the ultraviolet ring, where the densest VLA-measured column density contour is 4.6 $\times$ 10$^{20}$ cm$^{-2}$, R = 18~kpc, and V = 260 \kms. This contour overlaps with the UV ring at its inner edge, but no other ultraviolet emission is evident. At this location, $\Sigma_{gas,crit}$ is 5.7 \Msun~pc$^{-2}$, while we actually observe $\Sigma_{gas, obs}$ to be 5.3 \Msun~pc$^{-2}$. Though these values are more consistent, leading to the expectation of star formation throughout, we observe UV star formation only at the innermost edge of this high density contour. 

Finally, we also measure a spot where the local \HI peak coincides exactly with a clump of ultraviolet emission in the northwest corner of the ring, best seen again in the VLA data overlay in Figure~\ref{dss}. Inside the highest \HI contour, 4.6 $\times$ 10$^{20}$ cm$^{-2}$, R $\sim$ 21~kpc and again assuming that V = 260 \kms, $\Sigma_{gas, crit}$ = 4.9 \Msun~pc$^{-2}$. At this position, $\Sigma_{gas, obs}$ = 5.3 \Msun~pc$^{-2}$. In this case, where the star formation entirely coincides with an \HI peak, the presence of star formation is consistent with the threshold calculated from the Kennicutt relation.

For each of these three locations in the ring, we can predict the $\Sigma_{SFR}$ using the relation between gas surface density and star formation surface density observed by \citet{Kennicutt98} in a sample of 61 normal spiral galaxies with well-studied \HI, H$\alpha$, and CO properties. At the southernmost point of the UV ring, the observed gas surface density yields a $\Sigma_{SFR}$ of 1.8 $\times$ 10$^{-3}$ \Msun~year$^{-1}$~kpc$^{-2}$, while at the higher column density contours of the southeastern and northwestern edges of the ring, $\Sigma_{SFR}$ is predicted to be 2.6 $\times$ 10$^{-3}$ \Msun~year$^{-1}$~kpc$^{-2}$. Averaged over the entire \HI ring and assuming the physical area of our \HI model and the factor of 1.45 for heavier elements, $\Sigma_{gas, obs}$ = 1.20 \Msun~pc$^{-2}$, yielding a $\Sigma_{SFR}$ = 3.2 $\times$ 10$^{-4}$ \Msun~year$^{-1}$~kpc$^{-2}$. 

However, from our NUV measurement normalized by the area of the entire ultraviolet ring, we find a surface density of star formation of 1.8 $\times$ 10$^{-4}$ \Msun~year$^{-1}$~kpc$^{-2}$, which is roughly a factor of 2 lower than the SFR predicted by Kennicutt for the \HI ring-averaged case and a factor of 10 lower than the SFR predicted for the dense regions of the ring coincident with UV emission. 
It should be pointed out that the physical relation published in \citet{Kennicutt89} was derived to correlate gas density and massive star formation as traced by H$\alpha$, not star formation which is traced by ultraviolet emission; this may be responsible for some of the overestimate. However, on a global scale, the physical conditions of the ISM in the ring around ESO 381-47 
are consistent with the finding of \citet{Bigiel08} that a simple power law cannot correlate $\Sigma_{gas}$ and $\Sigma_{SFR}$ for all levels of star formation, and the star formation efficiency depends strongly on the radius \citep{Leroy08}, with outer disk star formation being the least efficient. The \HI content and star formation rate as measured by the UV of ESO 381-47 are consistent with the sub-Kennicutt outer disk star formation observed by these authors, though their work focuses on star formation within D$_{25}$, while the ring around ESO 381-47 lies outside of this radius.

\subsection{ESO 381-47 as XUV disk}

The gas-rich, very low surface brightness (LSB) disk of ESO 381-47 with bright UV spots is reminiscent of the \citet{Ferguson98} and GALEX results that many low surface brightness disk galaxies are unstable to star formation far beyond the bright optical body. In recent work by \citet{Thilker07}, recent star formation at large galactic radii in (almost always) extended \HI distributions was studied in a sample of nearby systems selected from the GALEX Atlas of Nearby Galaxies. These galaxies are classified as XUV-disks, and specifically Type 1 XUV-disks are labelled as such by the presence of structured, UV-bright emission located outside of the bright, main body of the galaxy as identified by a star formation surface density threshold of 3$\times$10$^{-4}$ \Msun~yr$^{-1}$ kpc$^{-2}$. In 75\% of Type 1 XUV-disks, an external interaction is evidently responsible for the external star formation, though the presence of the gas in the outer disks may or may not be related to the same interaction. \citet{Thilker07} conclude that XUV-disk star formation is most likely triggered by the arrival of an interacting system in cold gas that is already present but quiescent.

We can therefore draw a parallel between ESO 381-47 and Type 1 XUV disks. 
The innermost region of the galaxy is bright enough to be above the threshold ultraviolet surface density, though this UV emission is actually due to the presence of old stars (not recent star formation as in the Thilker sample). The structured, UV-bright, optically-faint emission of the ring lies outside of this innermost contour. Therefore, we suggest that ESO 381-47, which for a long time was called an elliptical galaxy, is an example of the extreme limit of the Type 1 XUV-disk population where there is no inner star forming disk and where we are able to actually search for the ``smoking gun" which triggered its recent star formation. 

Other galaxies in the XUV-disk catalog of \citet{Thilker07} do resemble ESO 381-47, though they are all later-type spirals. NGC 5055 exhibits a warped \HI disk with outer spiral star formation features, NGC 1512 has an inner bright star forming ring but also a bar, and NGC 2841 shows ring-like star formation just outside the optical disk with a similar \HI morphology. The similarity of ESO 381-47 to Type 1 XUV-disks, which often are gas-rich and owe their UV features to perturbing companions, lack of a bar, and the presence of a companion at the radius of the ring may each point to an external interaction as the specific trigger for the star forming ring. 



\subsection{The ring morphology}

The stellar characteristics and SFR of the ring do not match those of the stellar population at the center of ESO 381-47. S08 find an age of 12.9 Gyr for the central region (R$_{e}$/16; R$_{e}$=22$\as$), while the age within R$_{e}$/2 is 8.9 Gyr. 
From the standpoint of the age of the stellar populations within the central galaxy, ESO 381-47 is a normal early type system with its oldest 
stars at the very center, but it is surrounded by a ring of young stars, which have likely been formed in the last 350-700 Myr. 

The lack of a gradual but consistent change in age 
from the center of the galaxy to the outermost ring may indicate that it is unlikely for a single event to be responsible for the creation of the stars at the center of the galaxy as well as the propagation of star formation out to the radius of the ring. However, given the massive \HI reservoir and the spread in ages indicated by the scatter in the color-color plot in Figure~\ref{colorage}, it is possible that the ring has been quietly forming stars at very low levels for several Gyr. A single event which continuously forms stars as it propagates from the center of the galaxy to the outskirts -- such as the suggestion in \citet{vanDriel91} that \HI rings around S0 galaxies could be formed from a ``burning out" of the gas on the insides of the galaxy -- may then be the case for the current ring around ESO 381-47. 

The age of the stellar populations in the interior of the galaxy are consistent with a 10~Gyr-old stellar population, but the star formation in the ultraviolet ring was much more recent. 
The old stellar population does become slightly younger and slightly less metal-rich with distance from the center, so such a ``burn out" must have either created only very low levels of new stars over that time or have been stalled for roughly 8 Gyr before forming the most recent burst of stars. If this is the case, then the disconnect between the stellar populations instead suggests that the star formation at the center of ESO 381-47 and its surrounding ring were triggered by separate events. In this section, we use the evidence provided by \HI, UV, and optical observations to speculate on the origin of the ultraviolet ring.

\subsubsection{Bars}

Merely being gas-rich is not sufficient for an early type galaxy to create young stars in its outskirts. Van Driel \& van Woerden (1991) find that barred early type galaxies tend to have outer \HI rings whose specific kinematics are governed by the bar. However, a brief search for star forming rings -- like the one we see in ESO 381-47 -- among the 24 objects in the van Driel sample shows that such features are not common. GALEX observed 13 of these 24 gas-rich S0 systems; only two out of these 13 systems, NGC 1291 and possibly NGC 4262, exhibit UV-bright rings, and both systems are barred. NGC 2974 is another example of an early type galaxy with an \HI ring of mass 5.5 $\times$ 10$^{8}$ \Msun~\citep{Weijmans08}, roughly a factor of 10 less massive than ESO 381-47, and a UV-bright star forming ring whose kinematics are governed by a large scale bar \citep{Jeong07}. No bar is detected in ESO 381-47 (see Figure~\ref{rband}), and so either the behavior of the ring must be ruled by some other force or interaction, or the bar which set up the ring was a transient one. 

The non-axisymmetric presence of a bound, tidally interacting companion can mimic the effect of a bar by creating a ring \citep{Combes88}, though statistically it is more likely that a companion would destroy an existing ring, break a true ring into a pseudoring structure, or prevent it from occurring in the first place \citep{Buta96}. The tightly wound spiral arm structure that is visible in the optical data, coincident with the star forming ring, may support a scenario in which an orbiting companion (ESO 381-47A) created pseudorings from an existing ring structure and in doing so initiated a new burst of star formation. Stars visible in the $R$-band indicate that the outer disk was previously populated with older stars even before this most recent burst of young stars, though the ages of the stellar component by S08 were not calculated at such a large galactic radius. 

In addition, inflow of gas as well as the arrival of a companion can disrupt an existing bar \citep{Buta96}, but there are no signs of a recent inflow in the form of cold gas, young stars, or an AGN at the center of ESO 381-47. The prediction in such a case of bar dissolution is for up to half of the stars to fall onto counter-rotating orbits without the stabilizing bar in place and for the ring structure to be destroyed if the encounter was a violent one. This seems unlikely, though it is not possible to ascertain whether the stars are counter-rotating due to the face on nature of the galaxy and ring, but certainly the ring has not been destroyed by such a major encounter.

\subsubsection{Spiral density waves}

\citet{Salo99} and \citet{Rautiainen00} model rings in the absence of a bar with spiral density waves. \citet{Salo99} find that the ``spiral only" potential is sufficient to create outer, detached spiral arms and a large, nuclear ring the size of typical inner rings. \citet{Rautiainen00} go a step further to create a family of simulations where the disk dominates the rotation curve except at large radii, where it is dominated by a moderate halo component. In the hottest disk (Toomre Q=2.5) simulation, a main bar is never formed, but a system of rings does appear including a flocculent, outer ring structure which is separated from the inner regions after roughly 10~Gyr. 

Such a scenario may be consistent with the optical morphology of ESO 381-47, including the inner ring seen in Figure~\ref{rband}. It can also explain the ``wiggles" in the velocity field of the VLA-only data shown in Figure~\ref{velfields}, which indicate local deviations from circular velocity and are characteristic of the presence of a density wave (e.g. \citealt{Visser80}). The most recent star formation in the ring would not need a second ``trigger" in this case; the dynamical evolution of the spiral wave spans more than 10~Gyr, and it is possible that the wave reaching the inner edge of the \HI ring was sufficient to begin forming stars.


\subsubsection{Accretion}

Gas accretion can form \HI rings, as in the cases of the early type galaxies IC 2006 \citep{Schweizer89,Franx94} and Hoag's Object \citep{Schweizer87}. Optically, these systems appear very similar to ESO 381-47. In the E1 galaxy IC 2006, the \HI ring and inner ionized gas counter-rotate with respect the stars at the center, which is taken to be evidence that the gas was accreted more recently than the formation of the stars at the center. In Hoag's Object, an E0, the stars and gas ring rotate in the same direction but the complete regularity of the central object suggests that the \HI and young stars must have been due to a later event. As there is no evidence of a companion, accretion is suggested by \citet{Schweizer87} as an alternate explanation. 

No velocity field of Hoag's Object is presented, but in IC 2006, the remarkable regularity of the velocity field is very unlike the large scale warp that we see in ESO 381-47. This implies perhaps that if gas has been accreted in the case of ESO 381-47, it has had less time to settle. 

\subsubsection{Collisional ring scenario}

Yet another possibility for the creation of the star forming ring around ESO 381-47 is a collisional ring scenario, wherein a density wave, set up by the bullseye collision of a companion with a gas rich disk, propagates from the center and moves outward. The maximum integrated gas density being located at slightly larger radii than the ultraviolet ring on the eastern and western edges is consistent with such a picture. In such collisional ring systems, stars are born out of the gas at what would have then been the densest part of the \HI ring, leaving adjacent parts of the ring to then be the most dense regions after the gas turns into stars (see \citealt{Appleton96} for a comprehensive review of collisional ring galaxies). Rarefactions left in the wake of the density wave are also predicted for these types of systems, as seen in the $R$-band residual image in Figure~\ref{rband}, though these residuals may only indicate that the subtracted model was too simple. A radial color gradient is also expected for collisional ring systems, and though the $V-R$ azimuthally averaged color profile displays such a gradient (as shown in Figure~\ref{vband}), the errors are very large at the radius of the ring $\sim$45-50~$\as$.

According to the stellar ages from the Bruzual \& Charlot models, the recently formed stars in the ring are consistent with an age of 350 - 700 Myr. To trigger starbirth 350 (700) Myr ago at a radius of 17.4~kpc, the wave would need to have been moving at roughly 50 (25) \kms to reach the current location of the UV ring. The most famous example of this type of system is the Cartwheel galaxy; models of this system find a propagation velocity of $\sim$100 \kms for the outer ring density wave, while the gas ring is radially expanding at only $\sim$65 \kms \citep{Vorobyov03}. \citet{Higdon96} analyzes the star formation in the Cartwheel, finding that only in the outermost ring does the surface density of \HI exceed the Kennicutt threshold (calculated from the relation published in \citealt{Kennicutt89}), though on the whole the gas surface density is much higher than that in ESO 381-47 at 5-28 \Msun~pc$^{2}$. The outer ring of the Cartwheel also exhibits coincident H$\alpha$ and radio continuum emission, indicating the presence of massive star formation, which is not present in ESO 381-47. The \HI mass in the Cartwheel 
is 9.3 $\times$ 10$^{9}$ \Msun~ 
\citep{Higdon96}, similar to what we see for ESO 381-47. 

The increased star formation and propagation velocity relative to the system that we present here could be due to the smaller mass ratio between the Cartwheel and its most probable intruder of roughly 10:1 (compared to the approximate mass ratio in this case of 80:1 if ESO 381-47A was the intruder). ESO 381-43N and 2MFGC 10338, each larger than ESO 381-47A, are also viable candidates for the collision, as either could have perpetrated the collisional ring and still reached their current locations, within roughly 160~kpc in projected distance from ESO 381-47, in 700 Myr with reasonable velocities of $\sim$220 \kms in the plane of the sky.

The morphology, possible presence of a radial color gradient, and similarity of the expansion velocity between ESO 381-47 and the Cartwheel may indicate that the system presented here is an example of a collisional ring. However, ESO 381-47 exhibits lower levels of star formation relative to the Cartwheel, which may be a result of the mass ratio of the companion or possibly even a different collisional geometry, where the ``hit" was less direct than a classical bullseye but was still able to excite a density wave in the gas disk. 

\subsection{\HI history and the environment}

The presence of ESO 381-47 in a gas-rich group, especially coupled with the small companion galaxy within the ring interacting strongly with the S0, may provide the answer to how the star forming ring was created. It seems unlikely for $\sim 10^{10}$ \Msun~of \HI to have been associated with the tiny ESO 381-47A; most likely the gas disk already existed before its arrival, though 
it is possible that the galaxy has had an effect upon the current \HI structure, as suggested by the PV diagram in Figure~\ref{pvs2}. The \HI in ESO 381-43N is also consistent with a picture wherein this galaxy was the original perturber, as it is concentrated opposite the hypothetical direction of travel. A scenario in which ESO 381-43N, 2MFGC 10338, and/or ESO 381-47A was the intruding system, if a perturber was in fact the star forming trigger, are all consistent with the time since the onset of star formation in the ring.

As it appears that the gas disk around ESO 381-47 existed before the arrival of the small companion galaxy, another question to ask is how the \HI arrived in the first place. Accretion, either inflow from the IGM or accretion from a gas-rich companion, is a strong possibility for the current state of the \HI. Many gas disks are warped, with the warp beginning at the optical disk edge \citep{VanderKruit07}, which makes ESO 381-47 more typical in this regard, and \citet{VanderKruit07} attribute such warped gas to accretion at a time much later than the formation of the inner disk. Further, it is not uncommon for gas-rich early-types to exhibit a hole in the neutral gas component corresponding to the location of the bulge. Gas exists inside this \HI hole, but in ionized form; in fact, S08 detected ionized gas in the form of [OIII], [NII], and H$\alpha$ within R$_{e}$/16 of ESO 381-47, and all of these plus H$\beta$ within R$_{e}$/2, and the kinematics of the ionized gas are more consistent with the kinematics of the \HI than those of the stellar component. 

Another possibility is presented by \citet{Brook07}, who find that a merger of two gas-rich disks with prograde rotation and a 2:1 mass ratio can also produce a completely merged remnant with a ring galaxy morphology after 1.5 Gyr. Such a picture is not impossible for ESO 381-47, though the optical luminosity in the ring is much fainter than in the simulation after only 100 Myr since the burst of star formation. Also, there is no significant spheroidal component in the resulting galaxy, but the authors attribute this to the lack of spheroids in the initial merging disks. One further difference between this simulation and our data is that the star formation rate predicted 1~Gyr after the merger -- 10 \Msun~yr$^{-1}$ -- is a factor of 100 higher than the integrated star formation rate that we see in the ring around ESO 381-47. 
If such a scenario occurred in the past for ESO 381-47, then ESO 381-47A would not be involved in the triggering of the star forming ring, lest the galaxy be undergoing two mergers (one major and one minor) in rapid succession. 

Additionally, for collisional ring systems, \citet{Mapelli08} find that after the first 100-200 Myr, the ring actually continues to expand and fade, and after a Gyr, the disk becomes large and flat, resembling giant low surface brightness galaxies (GLSBs; e.g. \citealt{Bothun90, Sprayberry95}). The ring stage for these systems is very brief. There is no prediction of ongoing star formation after a Gyr. However, comparing the ring around ESO 381-47 to the LSB galaxies included in the analysis by \citet{Bigiel08} shows that it shares a similar location in $\Sigma_{gas}$/$\Sigma_{SFR}$ parameter space with these faint systems.

Finally, the existence of a coherent velocity structure which connects all five of the galaxies in this group may suggest that they were more physically connected at some point in their past.
One possibility is that the systems were all formed out of a single filament or tidal stream of \HIcom as especially the smaller systems ESO 381-43 and ESO 381-43N resemble tidal dwarf galaxies. Perhaps a more likely possibility is that the \HI around ESO 381-47 was warped by a merger of gas rich systems at early times, creating in the process two gas-rich tidal streams from which the other galaxies in the group were formed. In either scenario, the current ring of star formation would certainly be unrelated to the creation of the warp, and presumably the old, central stellar population was created either after such a gas-rich merger or the collapse of gas within a primordial stream.

\section{Conclusions}

ESO 381-47 is an early type galaxy surrounded by an \HI ring with a mass of 6.7 $\times$10$^{9}$ \Msun. The \HI ring is warped but well modeled by regular, circular rotation. Ionized gas is detected in the central galaxy, and its kinematics are consistent with the inward extrapolation of the kinematics of the \HI ring. Neither the ionized gas nor the \HI kinematics are consistent with the stellar kinematics at the center of ESO 381-47. 

Near the inner edge of the \HI ring, but at smaller radius than the peak of the \HI, is a ring of ultraviolet emission indicative of recent star formation at a rate of 1.8 $\times$ 10$^{-4}$ \Msun~yr$^{-1}$~kpc$^{-2}$. From \citet{Kennicutt89}, we estimate the threshold gas density necessary for star formation to occur, and we find that in some cases stars are forming where our measurement of the local \HI density is roughly consistent with this threshold, and in other cases stars seem to be forming in sub-threshold areas. Though this may be due to the fact that local gas densities may be underestimated due to beam smearing, finding low level star formation at low $\Sigma_{gas}$ that departs from the Kennicutt relation is a general result also found by \citet{Bigiel08} in the outer parts of disk galaxies.  

We estimate the mean formation timescale in the ring by comparing the UV-optical colors with \citet{BC03} stellar population synthesis models which include a burst and exponentially declining star formation with various timescales and metallicities, and we find that a mix of ages is most likely present in the ring. Comparing the $\mathrm{FUV-NUV}$ colors to the same models as well as a model of an instantaneous burst also help to constrain the time since the most recent burst, which appears to be 350-700 Myr. 

The UV-bright features of ESO 381-47 are similar to those seen by \citet{Thilker07} in work done to classify XUV-disks. In particular, this system resembles the Type 1 XUV-disks in that it exhibits UV-bright/optically faint structure outside of a defined threshold star formation surface density. This makes ESO 381-47 (for a long time assumed to be an E) a very early type example of such systems, which span a morphological range of early- to late-type spirals.

In this paper, we discuss several ways to create the ultraviolet ring of recent star formation around ESO 381-47:
\begin{itemize}
\item Rings can be created by resonances set up by bars in the central galaxy, but no bar is evident from our deep optical imaging. If this was the cause for the formation of the \HI ring and/or the ring seen in the ultraviolet (and faintly in the optical), then the bar has disappeared by the present day. The arrival of a companion as well as gas inflow can disrupt a bar.
\item Alternatively, a spiral density wave, could be responsible for the faint optical appearance and coincident, UV-enhanced ring of ESO 381-47. \citet{Rautiainen00} describe such a wave set up in a disk too hot to form a stellar bar which becomes unstable at the edges to tightly wound spiral arm formation. This explanation does not require a separate ``trigger" for the ring of recent star formation after the initial formation of the density wave, as the ring structure emerges after 10~Gyr of evolution.
\item Accretion of gas may also play a role in the creation of the \HI ring, as is suggested for galaxies with \HI warps \citep{VanderKruit07} and as has been suggested for systems similar to ESO 381-47, such as IC 2006 and Hoag's Object. If such accretion happened within the last 10$^{8}$ years, it could also be responsible for the young stars that we see in the ring.
\item A collisional ring system wherein the small companion that we call ESO 381-47A, which exhibits a stellar tidal tail pointing toward the center of the system, and/or one of the nearby companion galaxies ESO 381-43N and 2MFGC 10338, may have triggered a radially expanding density wave which created the young stars seen at the inner edge of the \HI ring. 
\item Finally, \citet{Brook07} suggest that two gas-rich disks in a prograde merger can produce a ring galaxy, but the timescale and star formation rate predictions from this model disagree with our observations of ESO 381-47.
\end{itemize}


In this paper, we identify several possibilities to account for the creation of the star forming ring around ESO 381-47 and observe this galaxy as it makes the transition from a normal, early type system to a (faintly) star forming one which may be on its way to becoming a giant low surface brightness galaxy.

\section{Acknowledgements}

Many thanks to George Heald and Gyula Jozsa for useful comments and helpful discussions. GALEX is a NASA funded Small Explorer Mission, and we acknowledge support from NASA grant \#NNG05GE39G. This work was supported in part by the National Science Foundation under grant \# 0607643 to Columbia University. We acknowledge the usage of the HyperLeda database (http://leda.univ-lyon1.fr). This research has made use of the NASA/IPAC Extragalactic Database (NED) which is operated by the Jet Propulsion Laboratory, California Institute of Technology, under contract with the National Aeronautics and Space Administration. Finally, we would like to acknowledge Sofie Serra for waiting to be born until the preparation of this paper reached its final days.


\label{lastpage}

\end{document}